\documentclass[manuscript]{acmart}

\AtBeginDocument{%
  \providecommand\BibTeX{{%
    \normalfont B\kern-0.5em{\scshape i\kern-0.25em b}\kern-0.8em\TeX}}}

\acmYear{2024}\copyrightyear{2024}
\setcopyright{rightsretained}
\acmConference[ACM FAccT '24]{ACM Conference on Fairness, Accountability, and Transparency}{June 3--6, 2024}{Rio de Janeiro, Brazil}
\acmBooktitle{ACM Conference on Fairness, Accountability, and Transparency (ACM FAccT '24), June 3--6, 2024, Rio de Janeiro, Brazil}
\acmDOI{10.1145/3630106.3658962}
\acmISBN{979-8-4007-0450-5/24/06}

\begin{document}

\everypar{\looseness=-1}
\linepenalty=1000


\title[Algorithmic Transparency and Participation through the Handoff Lens]{Algorithmic Transparency and Participation through the Handoff Lens: Lessons Learned from the U.S.\ Census Bureau’s Adoption of Differential Privacy}

\author{Amina A. Abdu}
\email{aabdu@umich.edu}
\affiliation{%
 \institution{University of Michigan}
 \streetaddress{}
 \city{Ann Arbor}
 \state{Michigan}
 \country{USA}
 \postcode{48109}
}
\author{Lauren M. Chambers}
\email{lauren@ischool.berkeley.edu}
\affiliation{%
 \institution{University of California, Berkeley}
 \streetaddress{}
 \city{Berkeley}
 \state{California}
 \country{USA}
 \postcode{94720}
}

\author{Deirdre K. Mulligan}
\email{dmulligan@berkeley.edu}
\affiliation{%
 \institution{University of California, Berkeley}
 \streetaddress{}
 \city{Berkeley}
 \state{California}
 \country{USA}
 \postcode{94720}
}

\author{Abigail Z. Jacobs}
\email{azjacobs@umich.edu}
\affiliation{%
 \institution{University of Michigan}
 \streetaddress{}
 \city{Ann Arbor}
 \state{Michigan}
 \country{USA}
 \postcode{48109}
}


\begin{abstract}
Emerging discussions on the responsible government use of algorithmic technologies propose transparency and public participation as key mechanisms for preserving accountability and trust. But in practice, the adoption and use of any technology shifts the social, organizational, and political context in which it is embedded. Therefore translating transparency and participation efforts into meaningful, effective accountability must take into account these shifts. We adopt two theoretical frames, Mulligan and Nissenbaum's \textit{handoff model} and Star and Griesemer's \textit{boundary objects}, to reveal such shifts during the U.S. Census Bureau’s adoption of differential privacy (DP) in its updated disclosure avoidance system (DAS) for the 2020 census. This update preserved (and arguably strengthened) the confidentiality protections that the Bureau is mandated to uphold, and the Bureau engaged in a range of activities to facilitate public understanding of and participation in the system design process. 
Using publicly available documents concerning the Census’ implementation of DP, this case study seeks to expand our understanding of how technical shifts implicate values, how such shifts can afford (or fail to afford) greater transparency and participation in system design, and the importance of localized expertise throughout. 
We present three lessons from this case study toward grounding understandings of algorithmic transparency and participation: (1) efforts towards transparency and participation in algorithmic governance must center values and policy decisions, not just technical design decisions; (2) the handoff model is a useful tool for revealing how such values may be cloaked beneath technical decisions; and (3) 
boundary objects alone cannot bridge distant communities without trusted experts traveling alongside to broker their adoption.
\end{abstract}



\maketitle

\section{INTRODUCTION}

{\let\thefootnote\relax\footnotetext{{The content herein represents the personal views of the authors and is not intended to reflect the views of the United States Government or any Federal agency.}}}
Recent work on values in technology attempts to understand how technological changes can produce fairer, more accountable, and more trustworthy systems \cite{selbst2019fairness, lepri2018fair, diakopoulos2016}. Transparency and participatory design are often proposed to advance these goals in both academic work and policy \cite{kaminski2020, ehsan2021expanding, coglianese2019transparency, biden2023executive, european2021laying}. However, scholars, critics, and advocates have raised complications and limitations of transparency and participation, particularly when adopted uncritically \cite{ananny2016, corbett2023, sloane2022, delgado2021stakeholder}.  
We argue that prevailing models of algorithmic transparency and participation stand to benefit from sociotechnical analysis of transparency and participation on-the-ground. As such, we focus on a single case study: the adoption of differential privacy in the 2020 Decennial Census. Differential privacy (DP) is a mathematical definition of privacy that leverages statistical uncertainty to provably limit leakage of any individual’s sensitive information-- in other words, random noise is added to data in order to reduce the possibility for re-identification \cite{dwork2006}. Under the leadership of Chief Scientist John Abowd, the Census Bureau implemented DP in its 2020 disclosure avoidance system (DAS), the mechanism used to manipulate census response data prior to publication to ensure confidentiality in accordance with Title 13.\footnote{The Census Bureau is obligated to protect the privacy and confidentiality of individual data according to Title 13 of the U.S.\ Code} Notably, the technical affordances of DP allowed the Bureau to make details of the DAS public for the first time without undermining confidentiality. The Bureau embraced this possibility, introducing many innovations in transparency and attempting to facilitate participation from a wide variety of experts and the public.

Despite these significant efforts, this newfound transparency did not produce the accountability and trust the Bureau hoped to engender \cite{hotz2022, ncai2021, kenny2021, ochoa2021}. The resulting controversy attracted the attention of critical scholars, who have attempted to adjudicate its history and implications \cite{boyd2022, bouk2021, nanayakkara2022}. We build upon this literature; by employing the \textit{handoff lens} \cite{mulligan2019}, we parse out how a seemingly technological transition - from its previous statistical disclosure (SDL) methods to DP - in fact altered the very function of disclosure avoidance as the Bureau’s methods, experts, and values were reconfigured.

Drawing on this case study, we demonstrate the utility of the handoff model for addressing calls from the critical algorithmic transparency literature to examine transparency in context. 
In particular, we show that the handoff model makes visible \textit{where} decisions about values are embedded within sociotechnical systems and identifies the configurations of human actors surrounding these decisions.

Our contribution is three-fold. First, we shed new insight on a case that has been of significant interest to both researchers and policymakers. We provide an account of the specific values decisions at the core of the adoption of DP and how participation was limited in these decisions despite significant efforts.
Second, we argue that values should be at the center of transparency and participation efforts and  demonstrate the utility of the handoff model for eliciting these values.
Finally, we highlight the need for understanding the role of experts in transparency and participation processes. While the literature has focused on developing documentation artifacts for transparency and participation, the census case highlights the insufficiency of artifacts alone to facilitate meaningful participation. Trusted individuals with requisite expertise must exist, within stakeholder communities, in order for such artifacts to be understood and adopted.

\section{RELATED WORK}\label{sec:related}

Calls to make technical systems more trustworthy and accountable point to \textit{transparency} and \textit{participation} as  key interventions. Such calls are common across academic \cite{kaur2022trustworthy,cooper2022accountability,wieringa2020account,diakopoulos2016,citron2007technological}, industry \cite{deloitte}, and governmental and NGO-based initiatives \cite{biden2023executive, european2021laying, OECD_2021, NIST_2023, ISO_2020}. Transparency efforts have called for visibility around choices about data, processes, or mathematical properties of algorithms \cite{kaminski2020}, e.g., through documentation of the development process
\cite{hutchinson2021towards,raji2020closing,diakopoulos2017algorithmic} or structured disclosures of data properties \cite{mitchell2019model, gebru2021datasheets, holland2020nutrition}.
Other efforts have been made to release code, make `inscrutable' algorithms interpretable or their decisions explainable \cite{ehsan2021expanding,loi2021transparency,citron2007technological}.  
\looseness=-1

Yet revealing such inner workings can fail to live up to promises of participation, contestability, or trust. Researchers have consistently cautioned that algorithmic transparency efforts may not  bring the benefits they promise \cite{ananny2016,kroll2017accountable,lepri2018fair, corbett2023,burrell2016}.
Making inscrutable algorithms explainable or readily available is insufficient without accounting for the social, organizational, and political power structures that shape their outcomes \cite{kroll2018fallacy,kroll2017accountable,ehsan2021expanding,lima2022conflict,bell2022}.
Intermediate objects for transparency efforts, such as structured disclosure documents, prioritize and shift ownership, effort, expertise and values \cite{wong2023seeing}. 
\looseness=-1
These efforts offer visibility without necessarily offering meaningful paths to accountability for public administration \cite{coglianese2019transparency,kaminski2020} or governance more broadly \cite{corbett2023,corbett2023power}.
Accountability therefore relies on enabling substantive participation and contestation
\cite{kaminski2020,corbett2023power,cooper2022accountability,kluttz2022shaping,vaccaro2019contestability}.
However, approaches to participation are complicated by the technical features of algorithmic systems and the broader social, political, and organizational contexts in which they operate \cite{levy2021algorithms, robinson2022voices, jasanoff2016}. Participation efforts -- whether participatory design activities, communicating with a wider range of stakeholders, or creating structured opportunities for feedback -- can then stand in for meaningful stakeholder empowerment. Indeed, critiques have shown how participation efforts can be exploitative and performative \cite{sloane2022,pierre2021getting,corbett2023power,delgado2021stakeholder}, where incentive structures can encourage ``participation-washing'' 
\cite{birhane2022power,costanza2020design,nedcooper2022systematic,pierre2021getting}. 
\looseness=-1

Contextual analysis is needed to support meaningful participation in (algorithmic) governance. 
Tools to articulate responsibility and decision-making can be disconnected from what practitioners need \cite{madaio2022assessing,veale2018fairness,sloane2022german}, but
articulating the decisions being made in the implementation of systems can better reveal where decision-making is happening in the first place \cite{kroll2021outlining,hutchinson2021towards,martin2019ethical,jacobs2021measurement,sloane2023introducingcontextual}. Transparency is needed not just of design choices, but also of the policy questions these choices often seek to answer \cite{mulligan2018saving}. 
Past literature has pointed to how the introduction or substitution of technologies can reconfigure values and social and political arrangements, and that understanding this context shift is key to responsible technology development \cite{selbst2019fairness,leveson2016engineering,mulligan2019, akrich1992}.
This missing lens on what happens when we substitute a new technology \cite{madaio2022assessing,selbst2019fairness}, and how that shifts organizational roles \cite{metcalf2019owning,wong2023seeing,sloane2022introducing} with technical and political impacts \citep[e.g.,][]{mulligan2019,jacobs2022hidden}, must then be considered to support meaningful participation. 
A fundamental challenge is to reveal how values and decisions change as a result of the introduction or substitution of a new technology: the handoff model that we discuss in the following section intervenes on this specific challenge. 
\looseness=-1

\section{THEORETICAL LENSES}\label{sec:lens}

\subsection{Handoff Model}
Mulligan \& Nissenbaum’s handoff model illuminates the values embedded in seemingly technical changes within a sociotechnical system, in particular when one part of a system is replaced with another \cite{mulligan2020}. By surfacing often overlooked reconfigurations of political and social relationships between technologies and people, the handoff model challenges narratives of linear technological progress.
\looseness=-1

The handoff model defines five elements of a system to interrogate in order to expose the changing values that it embodies:  the \textit{functions} of a system and sub-functions of its components (what does it do?); the \textit{components} that are involved, be they technical or human (what are its pieces?); the ways in which components \textit{engage with} or \textit{act on} one another (where do its pieces connect?); the \textit{modes of action} by which one component acts on or engages another (in what ways do its pieces connect?); and finally the \textit{trigger} that spurs the handoff (why did it change?). 
\looseness=-1

The handoff model has previously been adopted to understand shifting values in a variety of sociotechnical systems, including the adoption of design practices within organizations \cite{wong2020} and shifts in technology design.
For instance, Mulligan and Nissenbuam use the model to reveal how changes in access control for mobile phones (introducing passcodes and, later, biometric verification) shifted the roles of developers and users as well as the space of privacy threats \cite{mulligan2020}.
\looseness=-1

In this paper, we apply the handoff model to examine the shift from SDL to DP. The function is ostensibly the same (confidentiality) but the handoff reveals the implicitly value-laden rearrangements of organizations and design decisions. 

\subsection{Boundary objects}
Star and Griesemer introduce the concept of boundary objects as arrangements that allow groups to work together without consensus \cite{star1989}. Star theorizes boundary objects as having three important characteristics: (1) interpretive flexibility, (2) arising from information needs that are (3) weakly structured in common use but strongly structured in local contexts \cite{star2010}. Past work on boundary objects in organizations has, for instance, shown that these intermediate arrangements can serve as important tools for learning and communication across diverse stakeholders for difficult or heterogeneous domains. 

Past work demonstrates that the benefit of boundary objects also depends on their design, which can highlight gaps in understanding and values across groups or failures to serve different stakeholder needs \cite{carlile2002}. Further, the role of expertise is crucial: the design and use of boundary objects unintentionally (and intentionally) shapes stakeholder engagement \cite{kimble2010, sarkki2020}. Within the handoff model, boundary objects can help us to understand the changing relationships between different actants in a handoff. In particular, we examine how the Bureau shaped participation through the artifacts that it introduced to structure negotiations. By attending to these sites of negotiation within the Census Bureau's adoption of DP, our goal is to illuminate how differences in values and expertise affect sociotechnical handoffs with consequences for participation, trust and accountability.
\looseness=-1

\section{APPLYING THE THEORETICAL LENSES}\label{applying}
In this section, we employ the handoff model to highlight the reconfigurations which took place during the Census Bureau's change to a new disclosure avoidance system (DAS) as they adopted DP. Specifically, we analyze changes in the elements which relate directly to \textit{the DAS's primary function: protecting the confidentiality of census responses.} In addition, we consider the artifacts that the Bureau introduced as part of the DP implementation process. 
\looseness=-1

\subsection{Handoff Triggers: New tech, new threats, new hype}

From Census Bureau documents and communications \cite[e.g.,][]{abowd2018, abowd2020modernizing, abowd2023, abowd2021declaration}, 
we identify three triggers that spurred the handoff, i.e., the adoption of DP. 
The first trigger enabled the handoff: the \textit{development of DP} 
in 2006 presented a promising new method \cite{dwork2006}. Second, the Bureau became aware of mounting evidence that increased computational power and data access might lead to new \textit{threats to privacy and confidentiality}  \cite{bureau2023why, abowd2023, keller2023}. These threats focused on future, potentially dangerous threats via reconstruction and/or re-identification attacks \cite{dinur2003} as well as current, realizable threats. For instance, Latanya Sweeney showed in 2000 that almost 90 percent of respondents to the 1990 U.S.\ Census could be identified using zip code, birth date, and gender alone \cite{sweeney2000}. In response to potential future threats, the Bureau conducted experimental attacks on its own statistical releases. While the experimental details could not be made public, the Bureau claimed their attack demonstrated that the re-identification of census records was indeed a credible threat \cite{hawes2021, abowd2023}.
Third, the social environment served as another trigger: specifically the \textit{`tech for good' hype}. 
A heightened cultural interest in framing and `solving' policy problems using stylized computational methods preceded the Bureau's decision to adopt DP. This interest was evidenced in part by 
the rise of programs such as Code for America in 2009, 
the United States Digital Service in 2014, 
and the Mechanism Design for Social Good initiative in 2016 \cite{green2019}, 
as well as by technical communities' heightened attention to questions of fairness, accountability, transparency, and ethics of technology, from the FATML origins in 2014 to the establishment of ACM venues such as FAccT (previous FAT*) and AIES in 2018 \cite{abebe2020, barocas2014fatml}.
DP saw increased in tandem with this hype, promising to encode values such as transparency, accountability, and privacy into a mathematical formulation \cite{seeman2023framing, wu2013}. In such an environment, the adoption of a theoretical computer science methodology to bolster privacy protection in the national Census - and a methodology which would increase transparency into the Bureau's processes - aligned perfectly with prevailing trends.

\subsection{Handoff Components: Shifting experts, techniques, and data}

The sociotechnical ecosystem surrounding the handoff of the DAS has many \textit{actor-components}, many of which remained largely unchanged throughout the transition to DP. For instance, the stakeholders and users who depend upon Census data products and the external agencies and groups (demographers, community groups, etc.) with whom the Bureau collaborates remained relatively stable throughout the handoff. However, the handoff also introduced new components, shifting the experts and technologies involved in delivering the DAS's confidentiality function. Below we compared those DAS components before and after the shift from SDL to DP.

\looseness=-1

\subsubsection{Technical Methods}
A key shift in the DAS was the substitution of SDL tools with DP mechanisms, a new set of confidentiality-preserving tools built on a definition of privacy from theoretical computer science. 
Under the previous SDL methods, Bureau statisticians protected census response data through methods like suppressing and swapping individual records. Under DP, however, randomly generated ``noise'' is algorithmically added to census data to preserve confidentiality.
This transition introduced two new subcomponents of particular note. First, the tunable \textit{epsilon} ($\varepsilon$) parameter is a direct measure of privacy loss in DP, in which small $\varepsilon$ reflects low privacy loss (i.e., high privacy, low accuracy) while large $\varepsilon$ reflects high privacy loss (i.e., low privacy, high accuracy). Second is \textit{post-processing}, a new step added to the data pipeline under DP to further modify the confidential data after the injection of randomized noise. This step ensures all final data products are non-negative integers, in order to assuage human interpreters who might be confused or put off by, for instance, a table reporting -48.12 people residing within a particular geography. 

\subsubsection{Data Invariants}\label{invariants}
The Census's outputs have significant consequences: because certain statistics inform resource allocation, accurate representation of population is important to a number of stakeholders, including voting rights advocates, state and municipal governments, tribal leadership, and even disaster recovery and public health personnel \cite{ochoa2021, ncai2021, wright2022}. For the 2010 census and those prior, some particularly significant counts (such as total state populations) were held \textit{invariant} under the DAS;\footnote{Specifically, we know that ``total population, voting-age population, number of housing units, number of occupied housing units, and number and type of group quarters were all invariants at the block level in 2000 and 2010 Census publications'' \cite{abowd2018disclosure}. Blocks are the smallest unit of geography recorded by the Census \cite{abowd2023} In 2010, for instance, the Census divided the country into over 11 million blocks.} in other words, they were not manipulated from their value `as counted' \cite{nasem2020, ncsl21}. However, invariants are incompatible with traditional DP -- zero noise requires an infinite privacy budget -- meaning that any count held invariant complicates DP's confidentiality guarantees \cite{abowd2022topdown, nasem2020}. 
\footnote{We note that Bureau messaging regarding invariants is inconsistent. Officials have said in some instances that invariants can be reconciled with DP but that they ``eat[...] the privacy-loss budget'' \cite[p.~27]{nasem2020}, but in others that invariants ``fundamentally violate[...] the central promise of differentially private solutions to controlling disclosure risk''\cite[p.~32]{abowd2021declaration}. As such, it is unclear whether invariant values should or should not be considered within the scope of the DAS.} 
As a result, the 2020 DAS reduced the number of counts that would be held invariant, most notably no longer publishing the population of census blocks as counted.
Thus, the reported population - as well as demographic characteristics such as race and age - of all geographies smaller than a state would be altered with DP before publication. 

\subsubsection{Experts}
Finally, the adoption of DP thus meant the introduction of a new class of experts to the DAS: theoretical computer scientists, specifically those well-studied in DP formulations. As a result, computer scientists were slotted into DAS design processes, for instance: serving alongside social scientists, policy researchers, political advocates, and corporate leaders on DP working groups for two census oversight committees; and completing contractual work 
to directly assist in implementing DP for the 2020 Demographic and Housing Characteristics tabulations \cite{haney2021}.
\looseness=-1

\subsection{Handoff Modes: Abstraction and constrained expertise}\label{subsec:modes}

The handoff model pays particular attention to differences between \textit{modes} of interaction between components. Attending to how these modes shift with the introduction of DP reveals how underlying values shift as well. 
\looseness=-1

Some of the ways the Bureau manipulated response data remained fundamentally the same for the 2020 census, such as using imputation to compensate for non-responses \cite{cantwell2021}. However, the introduction of DP changes the mode in which the \textit{DAS acts on the Census response data} in order to protect respondents' confidentiality, shifting the roles of \textit{expert decision-making}. Under the Bureau's prior SDL methods, Bureau experts researched, debated, and determined the process for matching and swapping individual `microdata' records and for suppressing tables \cite{mckenna2018}. 
\looseness=-1
With the change to DP, however, the numerous swapping, rounding, or suppression decisions that a given DAS researcher might make regarding the particulars of disclosure avoidance are condensed down to fewer, more abstracted decisions - for instance, determining the value of the privacy budget $\varepsilon$ and the allocation of that privacy budget across the dataset.
\footnote{To be sure, there are still many fundamental choices that remain for designers of the DAS under DP. Researchers must draw block and tract boundaries, determine which data will be kept invariant, as well as which kinds of geographies will and will not include in the hierarchy known as the `spine,' to name a few \cite{abowd2023, ncsl21, cb2021}. Nevertheless, epsilon-DP greatly reduces the dimensionality of \textit{disclosure avoidance decisions} left to Bureau experts.}

The changes to the DAS also reconfigured the relevance of \textit{disciplinary expertise}. Many kinds of pre-DP census expertise (such as that of demographers and political advocates) was no longer sufficient to afford a confident understanding of, or even engagement with, how the DAS operates \cite{ncai2021, ochoa2021}. This is to say, the responsibility to effectively design privacy protections was displaced from statisticians and the DAS's prior experts onto computer scientists. This reorientation leads to not only a new balance of power across the landscape of Census experts, but also a fundamental shift in the rhetorical and epistemological configuration of the Census and the DAS \cite{boyd2022, nanayakkara2022}.

Finally, the primary mode of \textit{securing confidentiality} transformed under DP. Under prior SDL methods, revealing the details of DAS methods would entirely undermine their effectiveness by allowing a motivated attacker to reverse strategies and reconstruct the unprocessed data. Consequently, SDL methods provided confidentiality via maintaining secrecy and withholding details about the DAS from the public  -- an approach commonly referred to as ``security by obscurity.'' Thus the degree of confidentiality which SDL provided could only be determined by ex-post evaluation by internal bureau experts, and it was not future-proofed.
With DP, however, confidentiality becomes mathematically provable, ex-ante, under the  guarantees provided by statistical uncertainty. Secrecy and public withholding are no longer a concern. What's more, this reconfiguration also shifts how - and even if - experts evaluate confidentiality. While SDL required expert judgement in order to evaluate how much confidentiality protection a particular configuration of the DAS afforded, DP has displaced that judgement onto a statistical guarantee. Experts are no longer required to assess how effective the DAS is at preserving confidentiality; DP does it for them via epsilon ($\varepsilon$), the privacy budget.

\subsection{Handoff Function: Interrogating the \textit{how}}\label{function}

At first blush, it seems that the disclosure avoidance system’s primary \textit{function} remained the same before and after DP: preserving the confidentiality of census responses. 
However, the handoff model encourages a richer understanding of a system's function, including not only its  ``goals, purposes, or [...] values'' but also ``\textit{how} it does what it does, as a designer or engineer might explain it'' \cite[p.~6, original emphasis]{mulligan2020}.
The handoff lens ultimately reveals that DP, by shifting \textit{how} the function of confidentiality preservation was enacted, expanded the function of the DAS in many value-laden ways. In particular, DP (1) created new opportunities for transparency between the Bureau and interested publics; (2) allowed for formal, quantifiable validation of the privacy and confidentiality commitments actualized by the Bureau; and (3) replaced one form of expertise with another, precipitating the rise of theoretical computer science professionals and the decline of statisticians in the design, operation, evaluation of disclosure avoidance. We explore these shifting functions in greater detail throughout Section \ref{sec:results}.

\subsection{Transparency artifacts at the boundaries: Spaghetti at the wall}\label{sec:bo}

To seize the benefits of the transparency that DP allows, and to enable stakeholder participation in the DAS design, the Bureau created many new artifacts to facilitate public understanding and input of the DAS. First, the Bureau released an unprecedented degree of technical detail, sharing the DAS \textit{source code} via GitHub \cite{haase2021}. Realizing that code was not sufficient for providing transparency given stakeholder capacity, the Bureau released \textit{demonstration data} that would allow demographers and social scientists who use census products to interact with the new system in a way that was familiar to them. Ultimately releasing six sets of data between 2018 and 2021, these public datasets were the result of applying the Bureau’s 2020 DP algorithm to data from the 2010 Census.
\looseness=-1

The Bureau also engaged external experts in formal and informal \textit{co-design processes}.
Specifically the Bureau solicited written comments from data users via the Federal Register \cite{fedregister2018}; encouraged user feedback after publishing each round of demonstration data; hosted and participated in
workshops devoted to discussing the use of DP \cite{nasem2020, hdsr2019}; and held multiple consultations with tribal leadership \cite{consults2020}. 
Reaching beyond those experts, the Bureau provided an impressive array of \textit{educational resources} designed for more diverse stakeholders and the interested public about the new DAS. Blogs narrating the Bureau’s plans and progress as they worked to implement DP were authored by the Census’ chief scientist himself \cite{abowd2018, abowd2019, abowd2020modernizing}. To build up stakeholders' understanding of what DP is and why it is a worthwhile tool, the Bureau developed interactive Python Jupyter notebooks \cite{desai2022},
webinars \cite{webinar2023}, handbooks \cite{cb2021}, and videos \cite{minutephysics2019} designed for a lay audience. 
\looseness=-1

However, while the bureau created many artifacts and processes to bolster the public's understanding and participation in the DAS design, they withheld one artifact that external DP experts needed to evaluate the DAS. 
\textit{Noisy measurement files} are an interim data product which contain the census data after the application of DP, but before post-processing removes negative or non-integer values.  
These files were not originally released by the Bureau but became an object of great interest. 
In 2021, a group of over 50 researchers, technologists, and city officials wrote to the Bureau requesting publication of the noisy measurement file, arguing that the release of this noisy data would expedite evaluations of the downstream effects of DP while still adhering to Title 13 privacy requirements \cite{dwork2021}. Initially, the Bureau denied a FOIA request to release this data, citing concerns about confusing the public by revealing the existence of more than ``one `true' data set'' \cite[p.~15]{boyd2022}; researchers were further frustrated \cite{hullman2022}. Yet a year and a half later, in April 2023, the Bureau did release noisy measurements of 2010 demonstration data \cite{cb2023}. \label{noisy}
\looseness=-1

While the transparency artifacts listed in this section originated in the Bureau's interest in helping engage, educate, and involve diverse stakeholder groups; they ultimately ended up being sites of active negotiation within the handoff to DP. This reflects an implicit understanding within the Bureau of the need for boundary objects \cite{star1989} to bridge the many stakeholder groups they hoped to engage in the design of the DAS.
We do not intend to interrogate the degree to which individual artifacts were or were not successful in acting as boundary objects. 
Rather, we present and emphasize the \textit{volume and variety} of artifacts which the Bureau developed to facilitate stakeholder participation. Building upon this foundation, in Section \ref{sec:results} we will evaluate the artifacts' overall effectiveness in \textit{doing boundary work} \cite{langley2019boundary} -- that is, allowing a variety of stakeholders to participate in and negotiate the handoff from SDL to DP. 
\looseness=-1

\section{UNCOVERING THE STAKES OF THE HANDOFF}\label{sec:results}

Applying the handoff model to our case as outlined in section \ref{applying}, we surface several value-laden shifts in the Census Bureau's adoption and implementation of DP, as well as the participatory processes that the Bureau introduced to negotiate this transition. We summarize these findings in Table \ref{tab:handoff-app}.
\looseness=-1

\bgroup
\def\arraystretch{1.8}
\begin{table}[htbp]
    \centering
    \caption{Applying the handoff model to uncover shifting values and functions in the case of the 2020 Census}

    \begin{tabular}{p{0.24\linewidth}p{0.33\linewidth}p{0.33\linewidth}} 
        \hline 
         \textbf{The Census Bureau...}&  \textbf{Through the handoff lens}& \textbf{Conclusions about shifting values \& functionality}\\ \hline
         Switched from statistical disclosure methods (SDL) to differential privacy (DP)&  The function of confidentiality is preserved, but how confidentiality is operationalized has changed in response to triggers \looseness=-1 & Reveals the contested nature of \textit{confidentiality} (\S\ref{sec:confidentiality})\\\hline
         Solicited feedback about what use cases data users value $\bullet$ Reduced the total number of counts that are held invariant&  Changing the boundaries of the system through decisions about what is inside and outside the scope of confidentiality protections& Demonstrates the significance of \textit{data utility} (\S\ref{sec:utility}) as a function of the DAS $\bullet$ Concerns about access to political and economic resources are in tension with concerns about confidentiality\\\hline
         Introduced decision about the parameter epsilon as a locus of stakeholder participation& Functions (confidentiality and data utility) are now explicitly measured, formalized as a quantifiable trade-off $\bullet$ Experts evaluate and enact confidentiality through different modes, now with data-independent, ex-ante characterizations of privacy risk&Prioritizing \textit{formalized} (\S\ref{sec:formalism}) notions of privacy and accuracy re-scope those concepts and imply the existence of an optimal trade-off $\bullet$ Expert decisions about data manipulation (through e.g. swapping) are disintermediated by  DP implementation\\\hline
         Released significantly more information about the DAS (e.g., source code, demo data, blog posts)& Due to new DP methods, some forms of transparency no longer a threat to the DAS’ confidentiality function&\textit{Transparency} (\S\ref{sec:transparency}) emerges as a value of the DAS political process, but not clear that transparency can meet all of its second-order goals $\bullet$ Expert autonomy curtailed by  public scrutiny\\\hline
         Attempted to solicit and scaffold both expert and public participation& Introduction of new boundary objects as components in the DAS policy process
         &\textit{Participation} (\S\ref{sec:participation}) was ostensibly broadened, but with insufficient support by trusted experts\\ \hline
    \end{tabular}
    \label{tab:handoff-app}
\end{table}
\egroup

\subsection{Confidentiality is the tip of the iceberg} \label{sec:confidentiality}

Switching from earlier SDL methods to DP, the DAS maintained the same function of protecting Census respondents' confidentiality. The Bureau emphasized this change as a narrative of progress and increased effectiveness: the Bureau framed DP as a ``modern'' alternative to prior methods and ``a new, advanced, and far more powerful confidentiality protection system'' \cite{abowd2018}. 
\looseness=-1

The handoff model allows us to look beyond this narrative of linear progress to understand the larger social and political implications of the new DAS. While the DAS's \textit{function} remained focused on confidentiality, examining the changing \textit{components and modes} used to achieve this function reveals a more complicated story. First, \textbf{the notion of confidentiality is itself contested}. The Bureau's decision to pursue confidentiality through \textit{disclosure avoidance} is itself a value-laden choice, shaped by its interpretation of the confidentiality requirement outlined in Title 13. The handoff model demonstrates that changing one part of the DAS is not merely a modular replacement of one technical component with another. Instead, the adoption of DP changed the meaning of the system's core confidentiality function by shifting what harms the DAS was designed to protect against. (Indeed, different conceptualizations of confidentiality led to significant conflict around the Bureau's use of DP \cite{nanayakkara2022}.)
In particular, the turn to DP enables two distinct confidentiality functions: 1) empirical protections against external reconstruction of \text{individual} records and 2) because of DP's emphasis on future-proof theoretical guarantees, plausible deniability for the Bureau against any \textit{future harms}. The latter function is a shift from earlier versions of the DAS where these guarantees could not be rigorously formalized. This expansion of the confidentiality function aligns with the Bureau's interpretation of the Title 13 confidentiality mandate, targeting worst-case risk and insulating the Bureau from both present and future legal liability.
The choice to operationalize the DAS's confidentiality function using DP is an important upstream policy decision but, because of the Bureau's interpretation of their legal mandate, not one over which stakeholders outside of the Bureau had input. Moreover, beyond the contested meaning of confidentiality, the handoff lens reveals that \textbf{confidentiality was far from the only value implicated} by the shift to DP. 
\looseness=-1

\subsection{Data Utility}\label{sec:utility}

The DAS attempts to balance confidentiality with \textit{data utility}. The goal is to implement confidentiality protections that, in theory, allow a wide range of stakeholders to access and use census data, while ensuring that census takers trust these confidentiality protections enough to disclose their information. But in response to the Bureau's decision to update the DAS, many stakeholders expressed concerns that the noise added under DP would render Census data unusable for many use cases \cite{vanriper2021, wright2022}. 
\looseness=-1
Through the handoff lens, we note that the adoption of DP shifted other components of the DAS: in particular, in order to minimize privacy loss under DP, the Bureau reduced the number of published statistics and the number of counts that were held invariant (i.e., not affected by confidentiality protections; recall \S\ref{invariants}). The decision to report specific invariant statistics reflect policy decisions about what use cases are most important and where data utility should be preserved above confidentiality -- what statistics are understood to be essential for democratic representation, versus those which are malleable. Notably, the Bureau rejected a request from the National Congress of American Indians to keep state-level data for tribal areas invariant \cite{cbinvariants2020}. 
\looseness=-1

Because Census data are closely tied to the allocation of political and economic resources, decisions about data utility impact the pursuit of values like equity and justice. 
When the Bureau solicited feedback from data users in the Federal Register,
the solicitation and the responses revealed unspoken agreements about data access and utility for a wide range of applications, including state and local government, public health, anti-discrimination efforts, research, and education efforts. Stakeholders had differing epistemic perspectives about what makes data ``good enough'' to be useful \cite{boyd2022}. These ethical and epistemic questions underlie a number of important policy decisions about how the DAS should operationalize and prioritize data utility across different settings. Notably, our analysis highlights a key reframing around utility: the switch to DP and its focus on formalism (\S\ref{sec:formalism}) meant that \emph{utility} was largely operationalized as \emph{accuracy}, thus collapsing this epistemic debate.
\looseness=-1

\subsection{Formalism}\label{sec:formalism}
The shift to DP introduced a formal definition of both privacy and data utility. While  SDL methods could be formalized as a series of rules, they did not allow the Bureau to quantify the resulting confidentiality protections. The Bureau highlighted the advantages of formalism, citing provable and externally verifiable guarantees \cite{abowd2018} as well as precision in balancing competing values \cite{abowd2019}. 

An extensive literature on quantification in the history and sociology of science examines why the call to numbers has been so powerful, particularly in bureaucratic contexts \cite{porter1995, desrosieres1998, espeland1998, deringer2018}. Quantitative approaches to ethical questions promise to make political decisions visible \cite{sunstein2002, espeland1998} and facilitate debate in a common language \cite{desrosieres1998, deringer2018}, creating an avenue toward accountability by facilitating participation in democratic deliberation.
 \looseness=-1

Specifically, the Bureau's embrace of formalism allowed stakeholders to reason about previously hidden policy decisions and 
 made salient the balance and trade-off between privacy and accuracy (the latter often standing in for data utility \S\ref{sec:utility}). This is particularly salient in discussions about the privacy parameter $\varepsilon$, which promised to let the Bureau trade off between privacy protection and accuracy by operationalizing these complex values in an explicitly quantifiable way. 
The definition of DP does not dictate any particular value of $\varepsilon$, which must instead be tuned in accordance with the specific normative context of a particular data use case and a (deemed-acceptable) degree of privacy loss \cite{dwork2019}, i.e., ``$\varepsilon$ is a parameter chosen by policy'' \cite{dwork2006}. As a result, determining the value of $\varepsilon$ became a central focus of policy discussions surrounding the Bureau's adoption of DP. The choice of what to quantify reflects decisions about \textit{where to draw attention} \cite{espeland2007}. We argue that the Bureau's focus on formalism in the new DAS narrowed the scope of stakeholder participation by drawing attention to a single parameter of interest rather than a suite of decisions involved in the DAS handoff. 
Although there are many implementation decisions involved in DP, the formalization of the privacy-accuracy tradeoff in a single parameter focused attention on a single highly visible policy decision. 

 Stakeholders had previously largely ignored this tradeoff, effectively taking the Bureau’s prior statistical releases as ground truth \cite{boyd2022}. 
 \looseness=-1
Yet in describing the choice facing stakeholders as a quantifiable tradeoff between accuracy and privacy, the Bureau \textit{implied that an optimal choice was possible} \cite{bell2022, abowdschmutte2019}. This optimization discourse fit poorly onto the realities of the DAS. The Bureau acknowledged this challenge, noting that in order to get what the Data Stewardship Executive Policy considered a reasonable level of accuracy, they had to select a value of $\varepsilon$  that was ``far higher than those envisioned by the creators of differential privacy'' \cite[p.~3]{garfinkel2018}. Indeed, some critics posited that the choice of $\varepsilon$  created an impossible problem: ``There may not be an overlap between the values of $\varepsilon$  that are considered stringent enough for privacy purposes and high enough for redistricting purposes'' \cite[p.~2]{kenny2021}. In this way, we see how the constraints of a more-formalized DAS backed the Bureau into a rhetorical corner. 
\looseness=-1
 
The quantification literature cautions that numbers can hide policy decisions beneath a veneer of scientific objectivity, producing legitimacy in highly contested decision-making settings and, at times, foreclosing external intervention \cite{porter1995}. Moreoever, the choice to quantify privileges that which is easily measurable \cite{espeland1998}. In the context of the DAS, the Bureau called for ``pre-specified, objective criteria'' \cite[p.~2]{jarmin2023depth} to compare privacy methodologies. We argue that privileging formally quantifiable confidentiality guarantees led to a \textit{sociologically unintuitive conceptualization of privacy} \cite{seeman2023between},  and one which does not capture, e.g., notions of data privacy that are dependent on context or social relations \cite{viljoen2021relational, nissenbaum2004privacy}. Thus, the Bureau's decision to formalize privacy harms as reidentification risk--and data utility as (lack of) statistical uncertainty--in accordance with DP were not neutral, but reflect particular assumptions about the values at the center of the DAS. 
\looseness=-1

\subsection{Transparency}\label{sec:transparency}

The Bureau initially emphasized confidentiality, not transparency, as the benefit of the new DAS \cite{abowd2018}. However, the handoff lens reveals that the shift to DP also changed the DAS's capacity to support transparency. Because DP enabled technical details of the new DAS to be made public without compromising confidentiality, transparency emerged as a principal value of the political process surrounding the DAS. This shift opens up new possibilities for transparent relationships between Bureau researchers and the public.
\looseness=-1

A closer examination of this handoff reveals that the notion of ``transparency'' is in fact standing in for, and masking, many different values. In particular, there were many different ideas about the goal of transparency efforts, making it difficult for the Bureau to succeed in achieving each simultaneously. The Bureau engaged in many types of transparency, going beyond simplistic information disclosures and attempting to engage multiple audiences. Despite this, some stakeholders maintained that the Bureau was not being sufficiently transparent \cite{allis2020, ochoa2021, boyd2022}. Understanding the Bureau's various information releases as efforts to create boundary objects where competing notions of transparency were negotiated, we can unpack these transparency efforts and understand the many values and conflicts subsumed under the umbrella of transparency.
\looseness=-1

\subsubsection{Transparency for Data Utility}
Transparent privacy mechanisms can enable well-informed data users to make valid statistical inferences using privatized data by properly accounting for the uncertainty introduced by the privacy mechanism \cite{gong2022, williams2023}. Because of this, transparency in the DAS can make Census data more useful for statistical applications \cite{gong2022}. If the purpose of transparency is to enhance data utility through appropriate uncertainty quantification, transparency can be narrowly defined. In this case, only technical details are relevant objects of transparency, while disclosures about why a particular decision was made or who made a decision are outside of the scope of transparency for data utility. 

Yet, even under this narrowed scope, transparency can be complicated. While transparency under DP allowed the specifics of the algorithmic design, such as the value of $\varepsilon$, to become direct objects of public scrutiny and discussion, in the face of substantial uncertainty around what might constitute an appropriate value of $\varepsilon$, evaluations of the new DAS’s privacy protections and utility constraints varied substantially across experts \cite{schneider2021, hullman2022}.
Additionally, any data-dependent post-processing (for example, the Bureau's decision to enforce non-negative counts) undermines analysts' ability to estimate uncertainty \cite{hotz2022}. To address this challenge, a group of experts requested access to the noisy measurement files, as described in \S\ref{noisy}, which did not include the Bureau's post-processing steps. However, the Bureau did not initially release these files, preventing external experts from realizing the promised data utility benefits of the DAS's transparency.
\footnote{Further complicating the issue of transparency for data utility, many Census advocates argued that uncertainty caused by DP noise injection was minor compared to other sources in the Census's data collection and processing unrelated to confidentiality \cite{steed2022}. However, with some partial exceptions, these sources of uncertainty were not made transparent, undermining the transparency efforts and foreclosing comparison to DP uncertainty. This highlights the importance of considering transparency efforts and sociotechnical systems within their larger context.}
\looseness=-1

\subsubsection{Transparency for Trust}
Increased trust is often cited as a primary benefit of transparency efforts \cite[e.g.,][]{rawlins2008, jahansoozi2006}. We can see that the Bureau's decisions about what to make transparent-- and what to keep hidden-- were shaped by the importance of trust in census products. For example, the decision not to release the noisy measurement files (at odds with the pursuit of data utility through transparency as outlined in the previous subsection) was intended to preserve trust in census counts by hiding implausible counts produced by the original DP processing. However, by keeping the noisy measurement files hidden, external DP experts were not able to fully evaluate the Bureau's implementation \cite{boyd2022}, ultimately undermining the trust the Bureau had hoped to preserve. 
\looseness=-1

Outside of the noisy measurement files, the Bureau made many elements of the DAS visible during the transition to DP. Despite the Bureau's increased transparency, however, a number of key stakeholders expressed distrust in census data products during the DAS handoff. The National Congress of American Indians expressed concern that the 2020 census data would be ``inaccurate and unusable'' 
\cite[p.~3]{wright2022}; similarly, organizers working to increase participation in the census questioned ``why they should bother putting in all this effort if the end data are going to be so noisy'' \cite{nasem2020}.  
\looseness=-1

Schnackenberg and Tomlinson suggest that trustworthiness perceptions are enhanced through disclosure, clarity, and accuracy \cite{schnackenberg2016}. Because of limited ability to disclose all relevant information-- including previous details of disclosure avoidance systems, details of the reconstruction attack, and ground-truth data-- stakeholders could not evaluate the Census Bureau’s choices through the information disclosures and demonstration data. In the absence of this additional information, the complex technical details of the system, along with bugs in the demonstration data products caused by the post-processing system, damaged, rather than enhanced, trust \cite{bouk2021}.
\looseness=-1

 Importantly, the shift from secrecy to transparency about the perturbations of census data drew attention to data alterations and their implications that had gone unnoticed, or at least unexamined, by many stakeholders. Thus, transparency undermined trust not only in the Bureau’s implementation of DP, but also in the value of insights gained from previous census products \cite{boyd2022}. Freeman argues that when trust between stakeholders and agencies is low, negotiations over policy implementation take on an adversarial character under which transparency can become dangerous \cite{freeman1997}. By frontstaging the hidden work involved in the disclosure avoidance system, the Bureau revealed that decisions involved in its design were not merely sparing stakeholders mundane technical details but were in fact obscuring important policy choices. While the introduction of DP allowed the Bureau to make behind-the-scenes decision-making processes visible, this visibility exposed the slippage between the backstage and the frontstage of agency discretion – to the detriment of trust. 
\looseness=-1

\subsubsection{Transparency for Accountability}
Stakeholders' ability to interrogate the data and report on its limitations helped the Bureau identify what aspects of the DAS were limiting the utility of the data for different purposes. Allowing stakeholders to engage with the data during a Census workshop revealed the post-processing stage of the DAS was introducing ``unacceptable and problematic data biases and distortions'' \cite{abowd2020modernizing} and required structural changes. This insight demonstrates the value of the Census Bureau's transparency efforts in producing a more accountable DAS.
\looseness=-1

Yet, accountability was often limited because of a lack of transparency in what DP implementation decisions were feasible for the Census, or what limiting factors were effectively immutable. 
For instance, only the Bureau had access to the details of the previous SDL methods and the DP framework does not readily allow for comparisons to non-DP methods, making it challenging to assess critiques that did not agree with the DP formalization of privacy as a starting point \cite{kenny2022comment, hotz2022balancing}. Without knowledge of what policy levers were available to them, stakeholders were constrained in their ability to change the DAS. 
\looseness=-1

Additionally, accountability was further hampered by difficulties bridging different expert groups. The Bureau needed to communicate in expert language to the relevant theoretical computer science community to convey expertise and facilitate feedback. Yet, the technical jargon necessary to elicit solid feedback from that expert community yielded communications that alienated other expert stakeholders. A letter in July 2022 from the National Congress of American Indians specifically requested that the Bureau avoid the use of jargon and technical terms in their communications with tribal leadership, citing that prior tribal consultations were ``at far too high literacy levels for a lay audience and were therefore not meaningful consultation sessions'' \cite{wright2022}. While the Bureau recognized the importance of translating across varied stakeholder groups \cite{consults2020, abowd2020modernizing}, the challenge of doing so proved difficult to overcome and presented a persistent obstacle to accountability.
\looseness=-1

\subsection{Participation}\label{sec:participation}

The Bureau’s process for engaging stakeholders around the 2020 Census included a number of innovations to support both democratic and technocratic elements of agency policy-making \cite{mulligan2019}. As outlined in \S\ref{sec:transparency}, DP newly allowed transparency in the DAS, which in turn enabled a wider range of actors to be made aware of and participate in policy decisions embedded within the DAS.\footnote{Stakeholders have participated in and influenced prior iterations of the DAS. For example, ``data user dissatisfaction'' led the Bureau to pursue alternative methods to table suppression in the wake of the 1970 and 1980 Censuses \cite[p.~4-6]{bureau2023, mckenna2018}. However the DAS's reliance on ``security by obscurity'' meant that external stakeholders could not participate directly in many design decisions.} 
\looseness=-1

Increased technocratic participation became clear: during this shift, the Bureau brought in a range of experts and opened itself up to external expert review. These experts considered not only the technical details of DP and the DAS, but also provided input and review of the Bureau’s communications around the system.
\looseness=-1

More democratic participation was less clear. Such participation was mediated by Bureau's choices about who constituted a relevant public and how to communicate with them. While the Census Scientific Advisory Committee's DP working group applauded the Bureau for their efforts to include multiple perspectives, the committee also noted that it was difficult to assess what perspectives were not included and that many relevant stakeholders might not have the awareness, time, or energy to engage in policy decisions around the Census's implementation of DP \cite{breidt2020}.
\looseness=-1

Nevertheless, the Bureau introduced multiple innovations to facilitate democratic participation around complex technical artifacts. First, the Bureau moved beyond static notions of engagement like the traditional notice-and-comment process (i.e., the Bureau releases information at one moment in time, after which the public provides feedback). Instead, they introduced dynamic and ongoing engagement using a mix of videos, webinars and other educational materials coupled with listening sessions and discussions throughout development. Second, the Bureau published data artifacts produced by different policy choices, to scaffold better understanding of those policy choices. These data sets allowed stakeholders to interactively and intuitively explore the impact of potential implementation choices on their equities. Crucially, these demonstration products revealed a desire for boundary objects that would bridge between SDL methods and DP and allow users to interact with the new system in their own varied contexts. Through these artifacts, the Bureau attempted to surface implementation decisions that would be understood and shaped by multiple communities of practice. The Bureau ultimately did incorporate user feedback into the DAS design process in several cases. When the Bureau announced their final choice of $\varepsilon$, they emphasized that it was selected in response to user feedback from demonstration data analysis and that it was ``exponentially higher'' than the value of $\varepsilon$ proposed alongside earlier (expert-designed, expert-led) artifacts.  Yet, the choice to focus on a single parameter, rather than a range of policy decisions, limited where democratic participation was possible \cite{epsilon2021}.
\looseness=-1

\section{BEYOND THE CENSUS: LESSONS FOR TRANSPARENCY AND PARTICIPATION}\label{sec:discussion}

Technological changes in a system are not only technical: they reconfigure the social, political, and organizational contexts in which they occur. Understanding these reconfigurations is crucial for responsible, trustworthy, and accountable systems. 
\cite{mulligan2018saving,leveson2016engineering}.
We offer three lessons about how understanding these reconfigurations can enable meaningful governance, where  transparency and participation interventions would otherwise fall short.
\looseness=-1

\subsection{Lesson 1: \ \textit{The handoff lens is a critical tool for surfacing values}}

Applying of the handoff model in the census case enabled us to systematically untangle the daunting knot of actors, components, modes, functions, and values that were involved. 
In particular,  
the handoff model allows us to understand that the Census Bureau's shift to DP did not merely produce a functionally equivalent disclosure avoidance system (DAS), preserving the core function of confidentiality protection.
Instead, the shift toward DP was a deeper sociotechnical shift, reconfiguring the human and technical actors involved in the DAS and, ultimately, the values and forms of expertise embedded within this system. 
\looseness=-1
While working toward more accountable and trustworthy technology, it is critical to understand how the mere introduction of interventions-- such as new efforts towards transparency and participation-- may change the system in unintended ways. The handoff model can help us to identify these changes. For instance, the Census case reveals how the abandonment of `security by obscurity,' intended as a win for transparency, precluded transparency along certain dimensions. The handoff model makes space for critical and complicated invocations of transparency and participation (\S\ref{sec:related}), in realistic, on-the-ground contexts. We suggest the handoff model as a tool that can help researchers and technologists to systematically move lessons from critical algorithm studies into practice. 
\looseness=-1

\subsection{Lesson 2: \ \textit{Beware objects without experts}}

While the FAccT community has advocated for a range of artifacts as interventions toward transparency and participation \citep[e.g.,][]{gebru2021datasheets,mitchell2019model,holland2020nutrition}, these artifacts have largely been divorced from the contextual changes they introduce 
\cite{sloane2023introducingcontextual,mulligan2018saving}. 
In our case the Census Bureau invested significantly in such interventions towards
transparency and participation. Going beyond simplistic information disclosures, they created an impressive variety of boundary objects through which stakeholders could negotiate decisions about the DAS (\S\ref{sec:bo}). Further, the Bureau implemented many considerations that members of the FAccT community (and beyond) have long advocated for: toward \textit{explainability} through stakeholder education efforts, toward \textit{contextual} transparency through products like the demonstration data, and toward \textit{contestability} through ongoing dialogue and levers for change (namely, the value and allocation of the privacy budget). Despite the Bureau's enormous efforts, however, these boundary objects were only partially successful in facilitating meaningful participation and accountability, and in some cases they ultimately undermined trust. 
\looseness=-1

The mobilization of any given boundary object is dependent not only upon the object itself, but also upon the motivation and orientation of those \textit{brokers} that span and connect communities \cite{kimble2010}. The Bureau's boundary objects were in need of  \textbf{trusted local experts to carry them} across community divides.
Nurturing such experts is not a trivial task. Yet without them, the collaborative outcomes for which boundary objects are created in the first place might never come to fruition. Future work should explore in more detail what effective boundary object brokerage might look like in practice.
A too narrow focus on artifacts can overlook the processes needed to engage them. 
  
As the Census Bureau case demonstrates, \textbf{boundary objects cannot travel alone}.
The Bureau's focus on creating boundary objects, however innovative, was insufficient to build trust and comprehension among a diverse ecosystem of stakeholders. The epistemological and disciplinary chasms separating the communities which the Census was attempting to bridge
were just too wide \cite{boyd2022}. 
We encourage the FAccT community to think about the expertise needed to shepherd and use such boundary objects effectively in order to broker meaningful trust and participation.
\looseness=-1

\subsection{Lesson 3: \ \textit{Transparency and participation should center values and policy}}

Through our case study, we can expand theoretical critiques of transparency and participation to better understand tensions on the ground. We highlight complexity of actualizing transparency and participation in practice: despite efforts to solicit feedback over technical and design decisions, the Bureau faced criticism for not being sufficiently transparent or enabling sufficient participation. 
While the lessons from any one case is necessarily limited, we argue that a significant revelation from this case is that transparency efforts should not be purely about technical decisions, and that participation efforts should not be purely about design decisions. Rather, \textbf{both transparency and participation efforts should foreground decisions about values}. Importantly, providing transparency into technical decisions alone is not enough to reveal these values decisions. In fact, focusing on technical decisions can bound participation by making certain policy choices visible while neglecting 
others. In the Census Bureau's adoption of DP, for instance, a narrow focus on the privacy-loss parameter, $\varepsilon$, privileged the privacy-accuracy trade-off (and with narrow conceptualizations of both privacy and accuracy). Meanwhile other value-laden policy levers---including how confidentiality should be conceptualized and operationalized, what data should be within the scope of the DAS protections and what should be held invariant, and how the Bureau might advance values like equity and collective benefit \cite{CAREPrinciples}---were often less visible and therefore less accessible to participation. We argue that by prioritizing the visibility of values and policy on the same level, or even above, the visibility of technical details, the FAccT community can better leverage transparency and participation toward accountability and trust. 
\looseness=-1

\section{CONCLUSION}
The adoption of differential privacy by the U.S. Census Bureau marked a pivot in their practices around transparent and participatory algorithmic governance. The complex nature of this adoption, and its subsequent impacts revealed the ways in which \textit{handoffs} in algorithmic adoption in government must be mediated by different stakeholders with different levels of expertise, including via the use of carefully-designed boundary objects, to allow for meaningful participation. The lessons learned here apply more broadly to processes of algorithmic adoption, well-intentioned (and carefully planned) shifts towards transparency, and practices for successful handoffs in modern algorithmic governance.

\section{RESEARCH ETHICS AND SOCIAL IMPACT}
\subsection{Ethical concerns}
\subsubsection{Methods}
To the best of our understanding, there are no significant ethical concerns inherent to this work, as it is based in analysis of only publicly available documents. No interviews or sensitive data were collected for this paper. As such, no IRB approval was sought, as this work does not interface with human subjects.

\subsubsection{Fairness}
As we undertook our analysis, we took care to consider and portray the opinions of the various communities involved as fairly and equitably as possible, while understanding that some of these communities have been in active disagreement around the specifics of the 2020 DAS for years. Due to representation in the public archive and space limitations, we acknowledge that we were not able to represent every stakeholder viewpoint nor every notable event in the history of the Bureau's DP implementation.

\subsection{Positionality}
All four authors are U.S. citizens. While we are all thus implicated in matters of American legislative representation and voting rights, we all also reside in well-resourced regions which are not threatened by census undercounts or exclusion, nor by active infringements upon voting rights.

One author participated in processes around the 2020 DAS in real time; the other three became involved the project post-2020, and do not belong to any of the primary stakeholder groups that were most active in the census debate.
One author is trained in computer science, two in the mathematical and physical sciences, and one as a lawyer; these backgrounds informed our comprehension of and perspectives on the legal and technical processes at play.

\subsection{Adverse impact statement}
The primary adverse impact that this work could have would be playing into the hands of those who would weaponize the census for political gain. Given the heavily politicized nature of the census in general, and of the DP debate in particular, we cannot anticipate how or whether this work could be used to undermine the legitimacy of the census. Further, given the importance of the census for essential societal processes such as redistricting and resource allocation (which we address in our paper), we cannot dismiss the potential for such weaponization as inconsequential. 

Unfortunately, there is indeed precedent for such adverse impact. During the DAS development process, the Bureau faced direct political threats to its data products, the most serious of which arose in March 2021 when the state of Alabama sued the Department of Commerce and the Census Bureau in federal district court, alleging that by adopting DP the Bureau had “manipulated” and “intentionally skewed” the redistricting data that they provided to states \cite{AlabamavCommerce}. Furthermore, the coincidence of the decennial count with the 2020 presidential election, as well as the uncertainty around the Trump administration's proposal to include a citizenship question on the census, drew political attention to the count. In a time when political actors were searching for any chinks in governmental armor, a Federal agency which was public about internal sources of error became an easy target. Indeed, the Bureau has faced bipartisan scrutiny for the troubles made evident by the implementation of DP –  including allegations that DP was a Trump administration tactic attempting to ‘game’ federal funding allocations, and directly contradictory allegations that DP was a Democratic tactic to destabilize the Trump administration \cite[p.~32]{bouk2021}. Of course, such critiques undermine the ultimate role of the Bureau - to produce representative population counts - and further muddy the already-cloudy waters when it comes to identifying an appropriate implementation of DP.

Ultimately, we believe that the benefits that publishing our analysis might provide - hopefully, insights for both more effective algorithmic governance and more critical algorithmic scholarship - outweigh any potential risks for further weaponization.
\looseness=-1

\begin{acks}
We would like to thank Jeremy Seeman, the  participants at Yale ISP's 2023 Data (Re)Makes the World Conference and the 2023 Privacy Law Scholars Conference. This material is based upon work supported by the National Science Foundation Graduate Research Fellowship Program under Grant No. DGE 2146752. Any opinions, findings, and conclusions or recommendations expressed in this material are those of the author(s) and do not necessarily reflect the views of the National Science Foundation. Deirdre K. Mulligan is a Professor at UC Berkeley School of Information, a faculty director of the Berkeley Center for Law and Technology, and currently serving as Principal Deputy US Chief Technology Officer in the White House Office of Science and Technology Policy. The content herein represents the personal views of the authors and is not intended to reflect the views of the United States Government or any Federal agency.
\end{acks}

\bibliographystyle{ACM-Reference-Format}
\bibliography{facct}


\begin{thebibliography}{138}


\ifx \showCODEN    \undefined \def \showCODEN     #1{\unskip}     \fi
\ifx \showDOI      \undefined \def \showDOI       #1{#1}\fi
\ifx \showISBNx    \undefined \def \showISBNx     #1{\unskip}     \fi
\ifx \showISBNxiii \undefined \def \showISBNxiii  #1{\unskip}     \fi
\ifx \showISSN     \undefined \def \showISSN      #1{\unskip}     \fi
\ifx \showLCCN     \undefined \def \showLCCN      #1{\unskip}     \fi
\ifx \shownote     \undefined \def \shownote      #1{#1}          \fi
\ifx \showarticletitle \undefined \def \showarticletitle #1{#1}   \fi
\ifx \showURL      \undefined \def \showURL       {\relax}        \fi
\providecommand\bibfield[2]{#2}
\providecommand\bibinfo[2]{#2}
\providecommand\natexlab[1]{#1}
\providecommand\showeprint[2][]{arXiv:#2}

\bibitem[Ala(2021)]%
        {AlabamavCommerce}
 \bibinfo{year}{2021}\natexlab{}.
\newblock \bibinfo{booktitle}{\emph{Alabama v. {U}.{S}. {Dep}'t of {Commerce}}}.
\newblock \bibinfo{publisher}{546 F. Supp. 3d 1057 (M.D. Ala.)}.
\newblock


\bibitem[Abebe et~al\mbox{.}(2020)]%
        {abebe2020}
\bibfield{author}{\bibinfo{person}{Rediet Abebe}, \bibinfo{person}{Solon Barocas}, \bibinfo{person}{Jon Kleinberg}, \bibinfo{person}{Karen Levy}, \bibinfo{person}{Manish Raghavan}, {and} \bibinfo{person}{David~G. Robinson}.} \bibinfo{year}{2020}\natexlab{}.
\newblock \showarticletitle{Roles for computing in social change}. In \bibinfo{booktitle}{\emph{Proceedings of the 2020 {Conference} on {Fairness}, {Accountability}, and {Transparency}}} \emph{(\bibinfo{series}{{FAT}* '20})}. \bibinfo{publisher}{Association for Computing Machinery}, \bibinfo{address}{New York, NY, USA}, \bibinfo{pages}{252--260}.
\newblock
\showISBNx{978-1-4503-6936-7}
\urldef\tempurl%
\url{https://doi.org/10.1145/3351095.3372871}
\showDOI{\tempurl}


\bibitem[Abowd et~al\mbox{.}(2022)]%
        {abowd2022topdown}
\bibfield{author}{\bibinfo{person}{John Abowd}, \bibinfo{person}{Robert Ashmead}, \bibinfo{person}{Ryan Cumings-Menon}, \bibinfo{person}{Simson Garfinkel}, \bibinfo{person}{Micah Heineck}, \bibinfo{person}{Christine Heiss}, \bibinfo{person}{Robert Johns}, \bibinfo{person}{Daniel Kifer}, \bibinfo{person}{Philip Leclerc}, \bibinfo{person}{Ashwin Machanavajjhala}, \bibinfo{person}{Brett Moran}, \bibinfo{person}{William Sexton}, \bibinfo{person}{Matthew Spence}, {and} \bibinfo{person}{Pavel Zhuravlev}.} \bibinfo{year}{2022}\natexlab{}.
\newblock \showarticletitle{The 2020 {Census} {Disclosure} {Avoidance} {System} {TopDown} {Algorithm}}.
\newblock \bibinfo{journal}{\emph{Harvard Data Science Review}} \bibinfo{number}{Special Issue 2} (\bibinfo{date}{jun 24} \bibinfo{year}{2022}).
\newblock
\newblock
\shownote{https://hdsr.mitpress.mit.edu/pub/7evz361i}.


\bibitem[Abowd(2018a)]%
        {abowd2018disclosure}
\bibfield{author}{\bibinfo{person}{John~M. Abowd}.} \bibinfo{year}{2018}\natexlab{a}.
\newblock \bibinfo{title}{Disclosure {Avoidance} for {Block} {Level} {Data} and {Protection} of {Confidentiality} in {Public} {Tabulations}}.
\newblock
\newblock
\urldef\tempurl%
\url{https://www2.census.gov/cac/sac/meetings/2018-12/abowd-disclosure-avoidance.pdf}
\showURL{%
\tempurl}


\bibitem[Abowd(2018b)]%
        {abowd2018}
\bibfield{author}{\bibinfo{person}{John~M. Abowd}.} \bibinfo{year}{2018}\natexlab{b}.
\newblock \bibinfo{title}{Protecting the {Confidentiality} of {America}'s {Statistics}: {Adopting} {Modern} {Disclosure} {Avoidance} {Methods} at the {Census} {Bureau}}.
\newblock
\newblock
\urldef\tempurl%
\url{https://www.census.gov/newsroom/blogs/research-matters/2018/08/protecting_the_confi.html}
\showURL{%
\tempurl}
\newblock
\shownote{Section: Government}.


\bibitem[Abowd(2021)]%
        {abowd2021declaration}
\bibfield{author}{\bibinfo{person}{John~M. Abowd}.} \bibinfo{year}{2021}\natexlab{}.
\newblock \showarticletitle{Declaration of {John} {M}. {Abowd}}.
\newblock In \bibinfo{booktitle}{\emph{State of {Alabama} v. {U}.{S}. {Department} of {Commerce}}}.
\newblock
\urldef\tempurl%
\url{https://censusproject.files.wordpress.com/2021/04/2021.04.13-abowd-declaration-alabama-v.-commerce-ii-final-signed.pdf}
\showURL{%
\tempurl}


\bibitem[Abowd and Hawes(2023)]%
        {abowd2023}
\bibfield{author}{\bibinfo{person}{John~M. Abowd} {and} \bibinfo{person}{Michael~B. Hawes}.} \bibinfo{year}{2023}\natexlab{}.
\newblock \showarticletitle{Confidentiality {Protection} in the 2020 {US} {Census} of {Population} and {Housing}}.
\newblock \bibinfo{journal}{\emph{Annual Review of Statistics and Its Application}} \bibinfo{volume}{10}, \bibinfo{number}{1} (\bibinfo{date}{March} \bibinfo{year}{2023}), \bibinfo{pages}{119--144}.
\newblock
\showISSN{2326-8298, 2326-831X}
\urldef\tempurl%
\url{https://doi.org/10.1146/annurev-statistics-010422-034226}
\showDOI{\tempurl}


\bibitem[Abowd and Schmutte(2019)]%
        {abowdschmutte2019}
\bibfield{author}{\bibinfo{person}{John~M Abowd} {and} \bibinfo{person}{Ian~M Schmutte}.} \bibinfo{year}{2019}\natexlab{}.
\newblock \showarticletitle{An economic analysis of privacy protection and statistical accuracy as social choices}.
\newblock \bibinfo{journal}{\emph{American Economic Review}} \bibinfo{volume}{109}, \bibinfo{number}{1} (\bibinfo{year}{2019}), \bibinfo{pages}{171--202}.
\newblock


\bibitem[Abowd and Velkoff(2019)]%
        {abowd2019}
\bibfield{author}{\bibinfo{person}{John~M. Abowd} {and} \bibinfo{person}{Victoria~A. Velkoff}.} \bibinfo{year}{2019}\natexlab{}.
\newblock \showarticletitle{Balancing privacy and accuracy: New opportunity for disclosure avoidance analysis}.
\newblock \bibinfo{journal}{\emph{Census Blogs}} (\bibinfo{year}{2019}).
\newblock


\bibitem[Abowd and Velkoff(2020)]%
        {abowd2020modernizing}
\bibfield{author}{\bibinfo{person}{John~M. Abowd} {and} \bibinfo{person}{Victoria~A. Velkoff}.} \bibinfo{year}{2020}\natexlab{}.
\newblock \showarticletitle{Modernizing disclosure avoidance: What we’ve learned, where we are now}.
\newblock \bibinfo{journal}{\emph{Census Blogs}} (\bibinfo{year}{2020}).
\newblock


\bibitem[Akrich(1992)]%
        {akrich1992}
\bibfield{author}{\bibinfo{person}{Madeleine Akrich}.} \bibinfo{year}{1992}\natexlab{}.
\newblock \showarticletitle{The {De}-{Scription} of {Technical} {Objects}}.
\newblock In \bibinfo{booktitle}{\emph{Shaping {Technology} / {Building} {Society}: {Studies} in {Sociotechnical} {Change}}}, \bibfield{editor}{\bibinfo{person}{Wiebe~E. Bijker}, \bibinfo{person}{John Law}, \bibinfo{person}{Trevor Pinch}, {and} \bibinfo{person}{Rebecca Slayton}} (Eds.). \bibinfo{publisher}{MIT Press}, \bibinfo{address}{Cambridge, MA, USA}, \bibinfo{pages}{208}.
\newblock
\showISBNx{978-0-262-02338-2}


\bibitem[Allis(2020)]%
        {allis2020}
\bibfield{author}{\bibinfo{person}{Kevin Allis}.} \bibinfo{year}{2020}\natexlab{}.
\newblock \bibinfo{title}{[Letter from Kevin Allis to Steven D. Dillingham]}.
\newblock
\newblock
\urldef\tempurl%
\url{https://archive.ncai.org/policy-research-center/research-data/recommendations/NCAI_Letter_to_US_Census_Bureau_on_DAS_6_25_2020_FINAL_signed.pdf}
\showURL{%
\tempurl}


\bibitem[Ananny and Crawford(2016)]%
        {ananny2016}
\bibfield{author}{\bibinfo{person}{Mike Ananny} {and} \bibinfo{person}{Kate Crawford}.} \bibinfo{year}{2016}\natexlab{}.
\newblock \showarticletitle{Seeing without knowing: Limitations of the transparency ideal and its application to algorithmic accountability}.
\newblock \bibinfo{journal}{\emph{New Media \& Society}} \bibinfo{volume}{20}, \bibinfo{number}{3} (\bibinfo{year}{2016}), \bibinfo{pages}{973--989}.
\newblock


\bibitem[Barocas and Hardt(2014)]%
        {barocas2014fatml}
\bibfield{author}{\bibinfo{person}{Solon Barocas} {and} \bibinfo{person}{Moritz Hardt}.} \bibinfo{year}{2014}\natexlab{}.
\newblock \bibinfo{title}{Scope}.
\newblock
\newblock
\urldef\tempurl%
\url{https://www.fatml.org/schedule/2014/page/scope-2014}
\showURL{%
\tempurl}


\bibitem[Bell et~al\mbox{.}(2022)]%
        {bell2022}
\bibfield{author}{\bibinfo{person}{Andrew Bell}, \bibinfo{person}{Ian Solano-Kamaiko}, \bibinfo{person}{Oded Nov}, {and} \bibinfo{person}{Julia Stoyanovich}.} \bibinfo{year}{2022}\natexlab{}.
\newblock \showarticletitle{It's {Just} {Not} {That} {Simple}: {An} {Empirical} {Study} of the {Accuracy}-{Explainability} {Trade}-off in {Machine} {Learning} for {Public} {Policy}}. In \bibinfo{booktitle}{\emph{Proceedings of the 2022 {ACM} {Conference} on {Fairness}, {Accountability}, and {Transparency}}} \emph{(\bibinfo{series}{{FAccT} '22})}. \bibinfo{publisher}{Association for Computing Machinery}, \bibinfo{address}{New York, NY, USA}, \bibinfo{pages}{248--266}.
\newblock
\showISBNx{978-1-4503-9352-2}
\urldef\tempurl%
\url{https://doi.org/10.1145/3531146.3533090}
\showDOI{\tempurl}


\bibitem[Biden(2023)]%
        {biden2023executive}
\bibfield{author}{\bibinfo{person}{Joseph~R. Biden}.} \bibinfo{year}{2023}\natexlab{}.
\newblock \showarticletitle{Executive order on the safe, secure, and trustworthy development and use of artificial intelligence}.
\newblock  (\bibinfo{year}{2023}).
\newblock


\bibitem[Birhane et~al\mbox{.}(2022)]%
        {birhane2022power}
\bibfield{author}{\bibinfo{person}{Abeba Birhane}, \bibinfo{person}{William Isaac}, \bibinfo{person}{Vinodkumar Prabhakaran}, \bibinfo{person}{Mark Diaz}, \bibinfo{person}{Madeleine~Clare Elish}, \bibinfo{person}{Iason Gabriel}, {and} \bibinfo{person}{Shakir Mohamed}.} \bibinfo{year}{2022}\natexlab{}.
\newblock \showarticletitle{Power to the people? opportunities and challenges for participatory AI}.
\newblock \bibinfo{journal}{\emph{Equity and Access in Algorithms, Mechanisms, and Optimization}} (\bibinfo{year}{2022}), \bibinfo{pages}{1--8}.
\newblock


\bibitem[Bouk and boyd(2021)]%
        {bouk2021}
\bibfield{author}{\bibinfo{person}{Dan Bouk} {and} \bibinfo{person}{danah boyd}.} \bibinfo{year}{2021}\natexlab{}.
\newblock \bibinfo{title}{Democracy's Data Infrastructure}.
\newblock
\newblock
\urldef\tempurl%
\url{http://knightcolumbia.org/content/democracys-data-infrastructure}
\showURL{%
\tempurl}


\bibitem[boyd and Sarathy(2022)]%
        {boyd2022}
\bibfield{author}{\bibinfo{person}{danah boyd} {and} \bibinfo{person}{Jayshree Sarathy}.} \bibinfo{year}{2022}\natexlab{}.
\newblock \showarticletitle{Differential {Perspectives}: {Epistemic} {Disconnects} {Surrounding} the {U}.{S}. {Census} {Bureau}'s {Use} of {Differential} {Privacy}}.
\newblock \bibinfo{journal}{\emph{Harvard Data Science Review}} \bibinfo{number}{Special Issue 2} (\bibinfo{date}{June} \bibinfo{year}{2022}).
\newblock
\urldef\tempurl%
\url{https://doi.org/10.1162/99608f92.66882f0e}
\showDOI{\tempurl}


\bibitem[Breidt et~al\mbox{.}(2020)]%
        {breidt2020}
\bibfield{author}{\bibinfo{person}{Jay Breidt}, \bibinfo{person}{Deborah Balk}, \bibinfo{person}{John Czajka}, \bibinfo{person}{Kathy Pettit}, \bibinfo{person}{Allison Plyer}, \bibinfo{person}{Kunal Talwar}, \bibinfo{person}{Richelle Winkler}, {and} \bibinfo{person}{Joe Whitley}.} \bibinfo{year}{2020}\natexlab{}.
\newblock \bibinfo{title}{Differential Privacy Working Group Deliverables: Report of the CSAC Differential Privacy Working Group}.
\newblock
\newblock
\urldef\tempurl%
\url{https://www2.census.gov/cac/sac/differential-privacy-wg-deliverables.pdf}
\showURL{%
\tempurl}


\bibitem[Burrell(2016)]%
        {burrell2016}
\bibfield{author}{\bibinfo{person}{Jenna Burrell}.} \bibinfo{year}{2016}\natexlab{}.
\newblock \showarticletitle{How the machine ``thinks'': {Understanding} opacity in machine learning algorithms}.
\newblock \bibinfo{journal}{\emph{Big Data \& Society}} \bibinfo{volume}{3}, \bibinfo{number}{1} (\bibinfo{date}{June} \bibinfo{year}{2016}), \bibinfo{pages}{205395171562251}.
\newblock
\showISSN{2053-9517, 2053-9517}
\urldef\tempurl%
\url{https://doi.org/10.1177/2053951715622512}
\showDOI{\tempurl}


\bibitem[Cantwell(2021)]%
        {cantwell2021}
\bibfield{author}{\bibinfo{person}{Pat Cantwell}.} \bibinfo{year}{2021}\natexlab{}.
\newblock \bibinfo{title}{How {We} {Complete} the {Census} {When} {Households} or {Group} {Quarters} {Don}'t {Respond}}.
\newblock
\newblock
\urldef\tempurl%
\url{https://www.census.gov/newsroom/blogs/random-samplings/2021/04/imputation-when-households-or-group-quarters-dont-respond.html}
\showURL{%
\tempurl}
\newblock
\shownote{Section: Government}.


\bibitem[Carlile(2002)]%
        {carlile2002}
\bibfield{author}{\bibinfo{person}{Paul~R. Carlile}.} \bibinfo{year}{2002}\natexlab{}.
\newblock \showarticletitle{A pragmatic view of knowledge and boundaries: Boundary objects in new product development}.
\newblock \bibinfo{journal}{\emph{Organization Science}} \bibinfo{volume}{13}, \bibinfo{number}{4} (\bibinfo{year}{2002}), \bibinfo{pages}{442--455}.
\newblock


\bibitem[Carroll et~al\mbox{.}(2020)]%
        {CAREPrinciples}
\bibfield{author}{\bibinfo{person}{Stephanie~Russo Carroll}, \bibinfo{person}{Ibrahim Garba}, \bibinfo{person}{Oscar~L Figueroa-Rodr{\'\i}guez}, \bibinfo{person}{Jarita Holbrook}, \bibinfo{person}{Raymond Lovett}, \bibinfo{person}{Simeon Materechera}, \bibinfo{person}{Mark Parsons}, \bibinfo{person}{Kay Raseroka}, \bibinfo{person}{Desi Rodriguez-Lonebear}, \bibinfo{person}{Robyn Rowe}, {et~al\mbox{.}}} \bibinfo{year}{2020}\natexlab{}.
\newblock \showarticletitle{The CARE principles for indigenous data governance}.
\newblock \bibinfo{journal}{\emph{Data Science Journal}}  \bibinfo{volume}{19} (\bibinfo{year}{2020}), \bibinfo{pages}{43--43}.
\newblock


\bibitem[Citron(2007)]%
        {citron2007technological}
\bibfield{author}{\bibinfo{person}{Danielle~Keats Citron}.} \bibinfo{year}{2007}\natexlab{}.
\newblock \showarticletitle{Technological due process}.
\newblock \bibinfo{journal}{\emph{Wash. UL Rev.}}  \bibinfo{volume}{85} (\bibinfo{year}{2007}), \bibinfo{pages}{1249}.
\newblock


\bibitem[Coglianese and Lehr(2019)]%
        {coglianese2019transparency}
\bibfield{author}{\bibinfo{person}{Cary Coglianese} {and} \bibinfo{person}{David Lehr}.} \bibinfo{year}{2019}\natexlab{}.
\newblock \showarticletitle{Transparency and algorithmic governance}.
\newblock \bibinfo{journal}{\emph{Administrative law review}} \bibinfo{volume}{71}, \bibinfo{number}{1} (\bibinfo{year}{2019}), \bibinfo{pages}{1--56}.
\newblock


\bibitem[Commission(2021)]%
        {european2021laying}
\bibfield{author}{\bibinfo{person}{European Commission}.} \bibinfo{year}{2021}\natexlab{}.
\newblock \showarticletitle{Laying down harmonised rules on artificial intelligence (Artificial Intelligence Act) and amending certain union legislative acts}.
\newblock \bibinfo{journal}{\emph{Eur Comm}}  \bibinfo{volume}{106} (\bibinfo{year}{2021}), \bibinfo{pages}{1--108}.
\newblock


\bibitem[Cooper et~al\mbox{.}(2022b)]%
        {cooper2022accountability}
\bibfield{author}{\bibinfo{person}{A~Feder Cooper}, \bibinfo{person}{Emanuel Moss}, \bibinfo{person}{Benjamin Laufer}, {and} \bibinfo{person}{Helen Nissenbaum}.} \bibinfo{year}{2022}\natexlab{b}.
\newblock \showarticletitle{Accountability in an algorithmic society: relationality, responsibility, and robustness in machine learning}. In \bibinfo{booktitle}{\emph{Proceedings of the 2022 ACM Conference on Fairness, Accountability, and Transparency}}. \bibinfo{pages}{864--876}.
\newblock


\bibitem[Cooper et~al\mbox{.}(2022a)]%
        {nedcooper2022systematic}
\bibfield{author}{\bibinfo{person}{Ned Cooper}, \bibinfo{person}{Tiffanie Horne}, \bibinfo{person}{Gillian~R Hayes}, \bibinfo{person}{Courtney Heldreth}, \bibinfo{person}{Michal Lahav}, \bibinfo{person}{Jess Holbrook}, {and} \bibinfo{person}{Lauren Wilcox}.} \bibinfo{year}{2022}\natexlab{a}.
\newblock \showarticletitle{A systematic review and thematic analysis of community-collaborative approaches to computing research}. In \bibinfo{booktitle}{\emph{Proceedings of the 2022 CHI Conference on Human Factors in Computing Systems}}. \bibinfo{pages}{1--18}.
\newblock


\bibitem[Corbett and Denton(2023)]%
        {corbett2023}
\bibfield{author}{\bibinfo{person}{Eric Corbett} {and} \bibinfo{person}{Emily Denton}.} \bibinfo{year}{2023}\natexlab{}.
\newblock \showarticletitle{Interrogating the T in FAccT}. In \bibinfo{booktitle}{\emph{Proceedings of the 2023 ACM Conference on Fairness, Accountability, and Transparency}}. \bibinfo{pages}{1624--1634}.
\newblock


\bibitem[Corbett et~al\mbox{.}(2023)]%
        {corbett2023power}
\bibfield{author}{\bibinfo{person}{Eric Corbett}, \bibinfo{person}{Emily Denton}, {and} \bibinfo{person}{Sheena Erete}.} \bibinfo{year}{2023}\natexlab{}.
\newblock \showarticletitle{Power and Public Participation in AI}. In \bibinfo{booktitle}{\emph{Proceedings of the 3rd ACM Conference on Equity and Access in Algorithms, Mechanisms, and Optimization}}. \bibinfo{pages}{1--13}.
\newblock


\bibitem[Costanza-Chock(2020)]%
        {costanza2020design}
\bibfield{author}{\bibinfo{person}{Sasha Costanza-Chock}.} \bibinfo{year}{2020}\natexlab{}.
\newblock \bibinfo{booktitle}{\emph{Design justice: Community-led practices to build the worlds we need}}.
\newblock \bibinfo{publisher}{The MIT Press}.
\newblock


\bibitem[Delgado et~al\mbox{.}(2021)]%
        {delgado2021stakeholder}
\bibfield{author}{\bibinfo{person}{Fernando Delgado}, \bibinfo{person}{Stephen Yang}, \bibinfo{person}{Michael Madaio}, {and} \bibinfo{person}{Qian Yang}.} \bibinfo{year}{2021}\natexlab{}.
\newblock \showarticletitle{Stakeholder Participation in AI: Beyond" Add Diverse Stakeholders and Stir"}.
\newblock \bibinfo{journal}{\emph{arXiv preprint arXiv:2111.01122}} (\bibinfo{year}{2021}).
\newblock


\bibitem[Deloitte(2020)]%
        {deloitte}
\bibfield{author}{\bibinfo{person}{Deloitte}.} \bibinfo{year}{2020}\natexlab{}.
\newblock \bibinfo{title}{Trustworthy AI: Bridging the ethics gap surrounding AI}.
\newblock
\newblock
\urldef\tempurl%
\url{https://www2.deloitte.com/us/en/pages/deloitte-analytics/solutions/ethics-of-ai-framework.html}
\showURL{%
\tempurl}


\bibitem[Deringer(2018)]%
        {deringer2018}
\bibfield{author}{\bibinfo{person}{William Deringer}.} \bibinfo{year}{2018}\natexlab{}.
\newblock \bibinfo{booktitle}{\emph{Calculated values: Finance, politics, and the quantitative age}}.
\newblock \bibinfo{publisher}{Harvard University Press}.
\newblock


\bibitem[Desai(2019)]%
        {desai2022}
\bibfield{author}{\bibinfo{person}{Uma Desai}.} \bibinfo{year}{2019}\natexlab{}.
\newblock \bibinfo{title}{uscensusbureau/census-dp}.
\newblock
\newblock
\urldef\tempurl%
\url{https://github.com/uscensusbureau/census-dp}
\showURL{%
\tempurl}


\bibitem[Desrosi{\`e}res(1998)]%
        {desrosieres1998}
\bibfield{author}{\bibinfo{person}{Alain Desrosi{\`e}res}.} \bibinfo{year}{1998}\natexlab{}.
\newblock \bibinfo{booktitle}{\emph{The politics of large numbers: A history of statistical reasoning}}.
\newblock \bibinfo{publisher}{Harvard University Press}.
\newblock


\bibitem[Diakopoulos(2016)]%
        {diakopoulos2016}
\bibfield{author}{\bibinfo{person}{Nicholas Diakopoulos}.} \bibinfo{year}{2016}\natexlab{}.
\newblock \showarticletitle{Accountability in algorithmic decision making}.
\newblock \bibinfo{journal}{\emph{Commun. ACM}} \bibinfo{volume}{59}, \bibinfo{number}{2} (\bibinfo{year}{2016}), \bibinfo{pages}{56--62}.
\newblock


\bibitem[Diakopoulos and Koliska(2017)]%
        {diakopoulos2017algorithmic}
\bibfield{author}{\bibinfo{person}{Nicholas Diakopoulos} {and} \bibinfo{person}{Michael Koliska}.} \bibinfo{year}{2017}\natexlab{}.
\newblock \showarticletitle{Algorithmic transparency in the news media}.
\newblock \bibinfo{journal}{\emph{Digital journalism}} \bibinfo{volume}{5}, \bibinfo{number}{7} (\bibinfo{year}{2017}), \bibinfo{pages}{809--828}.
\newblock


\bibitem[Dinur and Nissim(2003)]%
        {dinur2003}
\bibfield{author}{\bibinfo{person}{Irit Dinur} {and} \bibinfo{person}{Kobbi Nissim}.} \bibinfo{year}{2003}\natexlab{}.
\newblock \showarticletitle{Revealing information while preserving privacy}. In \bibinfo{booktitle}{\emph{Proceedings of the twenty-second {ACM} {SIGMOD}-{SIGACT}-{SIGART} symposium on {Principles} of database systems}}. \bibinfo{publisher}{ACM}, \bibinfo{address}{San Diego California}, \bibinfo{pages}{202--210}.
\newblock
\showISBNx{978-1-58113-670-8}
\urldef\tempurl%
\url{https://doi.org/10.1145/773153.773173}
\showDOI{\tempurl}


\bibitem[Dwork et~al\mbox{.}(2021)]%
        {dwork2021}
\bibfield{author}{\bibinfo{person}{Cynthia Dwork}, \bibinfo{person}{Gary King}, \bibinfo{person}{Ruth Greenwood}, \bibinfo{person}{William~T. Adler}, {and} \bibinfo{person}{Joel Alvarez}.} \bibinfo{year}{2021}\natexlab{}.
\newblock \bibinfo{title}{Re: {Request} for release of "noisy measurements file" by {September} 30 along with redistricting data products}.
\newblock
\newblock
\urldef\tempurl%
\url{https://gking.harvard.edu/files/gking/files/2021.08.12_group_letter_to_abowd_re_noisy_measurements.pdf}
\showURL{%
\tempurl}


\bibitem[Dwork et~al\mbox{.}(2019)]%
        {dwork2019}
\bibfield{author}{\bibinfo{person}{Cynthia Dwork}, \bibinfo{person}{Nitin Kohli}, {and} \bibinfo{person}{Deirdre Mulligan}.} \bibinfo{year}{2019}\natexlab{}.
\newblock \showarticletitle{Differential {Privacy} in {Practice}: {Expose} your {Epsilons}!}
\newblock \bibinfo{journal}{\emph{Journal of Privacy and Confidentiality}} \bibinfo{volume}{9}, \bibinfo{number}{2} (\bibinfo{date}{Oct.} \bibinfo{year}{2019}).
\newblock
\showISSN{2575-8527}
\urldef\tempurl%
\url{https://doi.org/10.29012/jpc.689}
\showDOI{\tempurl}


\bibitem[Dwork et~al\mbox{.}(2006)]%
        {dwork2006}
\bibfield{author}{\bibinfo{person}{Cynthia Dwork}, \bibinfo{person}{Frank McSherry}, \bibinfo{person}{Kobbi Nissim}, {and} \bibinfo{person}{Adam Smith}.} \bibinfo{year}{2006}\natexlab{}.
\newblock \showarticletitle{Calibrating {Noise} to {Sensitivity} in {Private} {Data} {Analysis}}. In \bibinfo{booktitle}{\emph{Theory of {Cryptography}}} \emph{(\bibinfo{series}{Lecture {Notes} in {Computer} {Science}})}, \bibfield{editor}{\bibinfo{person}{Shai Halevi} {and} \bibinfo{person}{Tal Rabin}} (Eds.). \bibinfo{publisher}{Springer}, \bibinfo{address}{Berlin, Heidelberg}, \bibinfo{pages}{265--284}.
\newblock
\showISBNx{978-3-540-32732-5}
\urldef\tempurl%
\url{https://doi.org/10.1007/11681878_14}
\showDOI{\tempurl}


\bibitem[Ehsan et~al\mbox{.}(2021)]%
        {ehsan2021expanding}
\bibfield{author}{\bibinfo{person}{Upol Ehsan}, \bibinfo{person}{Q~Vera Liao}, \bibinfo{person}{Michael Muller}, \bibinfo{person}{Mark~O Riedl}, {and} \bibinfo{person}{Justin~D Weisz}.} \bibinfo{year}{2021}\natexlab{}.
\newblock \showarticletitle{Expanding explainability: Towards social transparency in ai systems}. In \bibinfo{booktitle}{\emph{Proceedings of the 2021 CHI Conference on Human Factors in Computing Systems}}. \bibinfo{pages}{1--19}.
\newblock


\bibitem[Eltinge et~al\mbox{.}(2019)]%
        {hdsr2019}
\bibfield{author}{\bibinfo{person}{John Eltinge}, \bibinfo{person}{Robert Sienkiewicz}, \bibinfo{person}{Michael~B. Hawes}, \bibinfo{person}{Quentin Brummet}, \bibinfo{person}{Edward Mulrow}, \bibinfo{person}{Kurt Wolter}, \bibinfo{person}{David Van~Riper}, \bibinfo{person}{Tracy Kugler}, \bibinfo{person}{Johnathan Schroeder}, \bibinfo{person}{JosÃ© Pacas}, \bibinfo{person}{Steven Ruggles}, \bibinfo{person}{Brian Asquith}, \bibinfo{person}{Brad Hershbien}, \bibinfo{person}{Shane Reed}, {and} \bibinfo{person}{Steve Yesiltepe}.} \bibinfo{year}{2019}\natexlab{}.
\newblock \bibinfo{title}{Differential {Privacy} for 2020 {US} {Census}}.
\newblock
\newblock
\urldef\tempurl%
\url{https://assets.pubpub.org/j2yr11kl/11587735061843.pdf}
\showURL{%
\tempurl}


\bibitem[Espeland and Stevens(1998)]%
        {espeland1998}
\bibfield{author}{\bibinfo{person}{Wendy~Nelson Espeland} {and} \bibinfo{person}{Mitchell~L. Stevens}.} \bibinfo{year}{1998}\natexlab{}.
\newblock \showarticletitle{Commensuration as a social process}.
\newblock \bibinfo{journal}{\emph{Annual Review of Sociology}} \bibinfo{volume}{24}, \bibinfo{number}{1} (\bibinfo{year}{1998}), \bibinfo{pages}{313--343}.
\newblock


\bibitem[Espeland and Vannebo(2007)]%
        {espeland2007}
\bibfield{author}{\bibinfo{person}{Wendy~Nelson Espeland} {and} \bibinfo{person}{Berit~Irene Vannebo}.} \bibinfo{year}{2007}\natexlab{}.
\newblock \showarticletitle{Accountability, quantification, and law}.
\newblock \bibinfo{journal}{\emph{Annu. Rev. Law Soc. Sci.}}  \bibinfo{volume}{3} (\bibinfo{year}{2007}), \bibinfo{pages}{21--43}.
\newblock


\bibitem[for Standardization(2020)]%
        {ISO_2020}
\bibfield{author}{\bibinfo{person}{International~Organization for Standardization}.} \bibinfo{year}{2020}\natexlab{}.
\newblock \bibinfo{title}{ISO/IEC TR 24028:2020 Overview of trustworthiness in artificial intelligence}.
\newblock
\newblock
\urldef\tempurl%
\url{https://www.iso.org/standard/77608.html}
\showURL{%
\tempurl}


\bibitem[Freeman(1997)]%
        {freeman1997}
\bibfield{author}{\bibinfo{person}{Jody Freeman}.} \bibinfo{year}{1997}\natexlab{}.
\newblock \showarticletitle{Collaborative governance in the administrative state}.
\newblock \bibinfo{journal}{\emph{UCLA L. Rev.}}  \bibinfo{volume}{45} (\bibinfo{year}{1997}), \bibinfo{pages}{1}.
\newblock


\bibitem[Garfinkel et~al\mbox{.}(2018)]%
        {garfinkel2018}
\bibfield{author}{\bibinfo{person}{Simson~L. Garfinkel}, \bibinfo{person}{John~M. Abowd}, {and} \bibinfo{person}{Sarah Powazek}.} \bibinfo{year}{2018}\natexlab{}.
\newblock \showarticletitle{Issues {Encountered} {Deploying} {Differential} {Privacy}}. In \bibinfo{booktitle}{\emph{Proceedings of the 2018 {Workshop} on {Privacy} in the {Electronic} {Society}}}. \bibinfo{publisher}{ACM}, \bibinfo{address}{Toronto Canada}, \bibinfo{pages}{133--137}.
\newblock
\showISBNx{978-1-4503-5989-4}
\urldef\tempurl%
\url{https://doi.org/10.1145/3267323.3268949}
\showDOI{\tempurl}


\bibitem[Gebru et~al\mbox{.}(2021)]%
        {gebru2021datasheets}
\bibfield{author}{\bibinfo{person}{Timnit Gebru}, \bibinfo{person}{Jamie Morgenstern}, \bibinfo{person}{Briana Vecchione}, \bibinfo{person}{Jennifer~Wortman Vaughan}, \bibinfo{person}{Hanna Wallach}, \bibinfo{person}{Hal~Daumé III}, {and} \bibinfo{person}{Kate Crawford}.} \bibinfo{year}{2021}\natexlab{}.
\newblock \showarticletitle{Datasheets for datasets}.
\newblock \bibinfo{journal}{\emph{Commun. ACM}} \bibinfo{volume}{64}, \bibinfo{number}{12} (\bibinfo{date}{Nov.} \bibinfo{year}{2021}), \bibinfo{pages}{86--92}.
\newblock
\showISSN{0001-0782}
\urldef\tempurl%
\url{https://doi.org/10.1145/3458723}
\showDOI{\tempurl}


\bibitem[Gong(2022)]%
        {gong2022}
\bibfield{author}{\bibinfo{person}{Ruobin Gong}.} \bibinfo{year}{2022}\natexlab{}.
\newblock \showarticletitle{Transparent privacy is principled privacy}.
\newblock \bibinfo{journal}{\emph{Harvard Data Science Review}} \bibinfo{number}{Special Issue 2} (\bibinfo{date}{June} \bibinfo{year}{2022}).
\newblock
\urldef\tempurl%
\url{https://doi.org/10.1162/99608f92.b5d3faaa}
\showDOI{\tempurl}


\bibitem[Green(2019)]%
        {green2019}
\bibfield{author}{\bibinfo{person}{Ben Green}.} \bibinfo{year}{2019}\natexlab{}.
\newblock \showarticletitle{"Good" isn't good enough}. In \bibinfo{booktitle}{\emph{Conference and Workshop on Neural Information Processing Systems}} \emph{(\bibinfo{series}{AI for Social Good Workshop})}. \bibinfo{address}{Vancouver}.
\newblock
\urldef\tempurl%
\url{https://aiforsocialgood.github.io/neurips2019/accepted/track3/pdfs/67_aisg_neurips2019.pdf}
\showURL{%
\tempurl}


\bibitem[Haase(2021)]%
        {haase2021}
\bibfield{author}{\bibinfo{person}{Kenneth Haase}.} \bibinfo{year}{2021}\natexlab{}.
\newblock \bibinfo{title}{uscensusbureau/{DAS}\string\_2020\string\_Redistricting\string\_Production\string\_Code}.
\newblock
\newblock
\urldef\tempurl%
\url{https://github.com/uscensusbureau/DAS_2020_Redistricting_Production_Code}
\showURL{%
\tempurl}


\bibitem[Haney et~al\mbox{.}(2021)]%
        {haney2021}
\bibfield{author}{\bibinfo{person}{Sam Haney}, \bibinfo{person}{William Sexton}, \bibinfo{person}{Ashwin Machanavajjhala}, \bibinfo{person}{Michael Hay}, {and} \bibinfo{person}{Gerome Miklau}.} \bibinfo{year}{2021}\natexlab{}.
\newblock \bibinfo{title}{Differentially {Private} {Algorithms} for 2020 {Census} {Detailed} {DHC} {Race} \& {Ethnicity}}.
\newblock
\newblock
\urldef\tempurl%
\url{https://doi.org/10.48550/arXiv.2107.10659}
\showDOI{\tempurl}
\newblock
\shownote{arXiv:2107.10659 [cs, stat]}.


\bibitem[Hawes(2021)]%
        {hawes2021}
\bibfield{author}{\bibinfo{person}{Michael Hawes}.} \bibinfo{year}{2021}\natexlab{}.
\newblock \bibinfo{title}{The {Census} {Bureau}'s {Simulated} {Reconstruction}-{Abetted} {Re}-identification {Attack} on the 2010 {Census}}.
\newblock
\newblock
\urldef\tempurl%
\url{https://www.census.gov/data/academy/webinars/2021/disclosure-avoidance-series/simulated-reconstruction-abetted-re-identification-attack-on-the-2010-census.html}
\showURL{%
\tempurl}
\newblock
\shownote{Section: Government}.


\bibitem[Holland et~al\mbox{.}(2020)]%
        {holland2020nutrition}
\bibfield{author}{\bibinfo{person}{Sarah Holland}, \bibinfo{person}{Ahmed Hosny}, \bibinfo{person}{Sarah Newman}, \bibinfo{person}{Joshua Joseph}, {and} \bibinfo{person}{Kasia Chmielinski}.} \bibinfo{year}{2020}\natexlab{}.
\newblock \showarticletitle{The {Dataset} {Nutrition} {Label}: {A} {Framework} to {Drive} {Higher} {Data} {Quality} {Standards}}.
\newblock In \bibinfo{booktitle}{\emph{Data {Protection} and {Democracy}}}, \bibfield{editor}{\bibinfo{person}{Dara Hallinan}, \bibinfo{person}{Ronald Leenes}, \bibinfo{person}{Serge Gutwirth}, {and} \bibinfo{person}{Paul~De Hert}} (Eds.). \bibinfo{series}{Data {Protection} and {Privacy}}, Vol.~\bibinfo{volume}{12}. \bibinfo{publisher}{Bloomsbury Publishing}, \bibinfo{pages}{1--26}.
\newblock
\showISBNx{978-1-5099-3275-7}
\newblock
\shownote{Google-Books-ID: F2HRDwAAQBAJ}.


\bibitem[Hotz et~al\mbox{.}(2022)]%
        {hotz2022balancing}
\bibfield{author}{\bibinfo{person}{V~Joseph Hotz}, \bibinfo{person}{Christopher~R Bollinger}, \bibinfo{person}{Tatiana Komarova}, \bibinfo{person}{Charles~F Manski}, \bibinfo{person}{Robert~A Moffitt}, \bibinfo{person}{Denis Nekipelov}, \bibinfo{person}{Aaron Sojourner}, {and} \bibinfo{person}{Bruce~D Spencer}.} \bibinfo{year}{2022}\natexlab{}.
\newblock \showarticletitle{Balancing data privacy and usability in the federal statistical system}.
\newblock \bibinfo{journal}{\emph{Proceedings of the National Academy of Sciences}} \bibinfo{volume}{119}, \bibinfo{number}{31} (\bibinfo{year}{2022}), \bibinfo{pages}{e2104906119}.
\newblock


\bibitem[Hotz and Salvo(2022)]%
        {hotz2022}
\bibfield{author}{\bibinfo{person}{V.~Joseph Hotz} {and} \bibinfo{person}{Joseph Salvo}.} \bibinfo{year}{2022}\natexlab{}.
\newblock \showarticletitle{A {Chronicle} of the {Application} of {Differential} {Privacy} to the 2020 {Census}}.
\newblock \bibinfo{journal}{\emph{Harvard Data Science Review}} \bibinfo{number}{Special Issue 2} (\bibinfo{date}{June} \bibinfo{year}{2022}).
\newblock
\newblock
\shownote{https://hdsr.mitpress.mit.edu/pub/ql9z7ehf}.


\bibitem[Hullman(2022)]%
        {hullman2022}
\bibfield{author}{\bibinfo{person}{Jessica Hullman}.} \bibinfo{year}{2022}\natexlab{}.
\newblock \bibinfo{title}{Show me the noisy numbers! (or not)}.
\newblock
\newblock
\urldef\tempurl%
\url{https://statmodeling.stat.columbia.edu/2022/12/28/show-me-the-noisy-numbers-or-not/}
\showURL{%
\tempurl}


\bibitem[Hutchinson et~al\mbox{.}(2021)]%
        {hutchinson2021towards}
\bibfield{author}{\bibinfo{person}{Ben Hutchinson}, \bibinfo{person}{Andrew Smart}, \bibinfo{person}{Alex Hanna}, \bibinfo{person}{Emily Denton}, \bibinfo{person}{Christina Greer}, \bibinfo{person}{Oddur Kjartansson}, \bibinfo{person}{Parker Barnes}, {and} \bibinfo{person}{Margaret Mitchell}.} \bibinfo{year}{2021}\natexlab{}.
\newblock \showarticletitle{Towards accountability for machine learning datasets: Practices from software engineering and infrastructure}. In \bibinfo{booktitle}{\emph{Proceedings of the 2021 ACM Conference on Fairness, Accountability, and Transparency}}. \bibinfo{pages}{560--575}.
\newblock


\bibitem[Jacobs(2021)]%
        {jacobs2021measurement}
\bibfield{author}{\bibinfo{person}{Abigail~Z Jacobs}.} \bibinfo{year}{2021}\natexlab{}.
\newblock \showarticletitle{Measurement as governance in and for responsible AI}.
\newblock \bibinfo{journal}{\emph{arXiv preprint arXiv:2109.05658}} (\bibinfo{year}{2021}).
\newblock


\bibitem[Jacobs and Mulligan(2022)]%
        {jacobs2022hidden}
\bibfield{author}{\bibinfo{person}{Abigail~Z Jacobs} {and} \bibinfo{person}{Deirdre~K Mulligan}.} \bibinfo{year}{2022}\natexlab{}.
\newblock \showarticletitle{The Hidden Governance in AI}.
\newblock \bibinfo{journal}{\emph{The Regulatory Review}} (\bibinfo{date}{July} \bibinfo{year}{2022}).
\newblock


\bibitem[Jahansoozi(2006)]%
        {jahansoozi2006}
\bibfield{author}{\bibinfo{person}{Julia Jahansoozi}.} \bibinfo{year}{2006}\natexlab{}.
\newblock \showarticletitle{Organization-stakeholder relationships: exploring trust and transparency}.
\newblock \bibinfo{journal}{\emph{Journal of management development}} \bibinfo{volume}{25}, \bibinfo{number}{10} (\bibinfo{year}{2006}), \bibinfo{pages}{942--955}.
\newblock


\bibitem[Jarmin et~al\mbox{.}(2023)]%
        {jarmin2023depth}
\bibfield{author}{\bibinfo{person}{Ron~S Jarmin}, \bibinfo{person}{John~M Abowd}, \bibinfo{person}{Robert Ashmead}, \bibinfo{person}{Ryan Cumings-Menon}, \bibinfo{person}{Nathan Goldschlag}, \bibinfo{person}{Michael~B Hawes}, \bibinfo{person}{Sallie~Ann Keller}, \bibinfo{person}{Daniel Kifer}, \bibinfo{person}{Philip Leclerc}, \bibinfo{person}{Jerome~P Reiter}, {et~al\mbox{.}}} \bibinfo{year}{2023}\natexlab{}.
\newblock \showarticletitle{An in-depth examination of requirements for disclosure risk assessment}.
\newblock \bibinfo{journal}{\emph{Proceedings of the National Academy of Sciences}} \bibinfo{volume}{120}, \bibinfo{number}{43} (\bibinfo{year}{2023}), \bibinfo{pages}{e2220558120}.
\newblock


\bibitem[Jasanoff(2016)]%
        {jasanoff2016}
\bibfield{author}{\bibinfo{person}{Sheila Jasanoff}.} \bibinfo{year}{2016}\natexlab{}.
\newblock \showarticletitle{Reclaiming the {Future}}.
\newblock In \bibinfo{booktitle}{\emph{The {Ethics} of {Invention}: {Technology} and the {Human} {Future}}}. \bibinfo{publisher}{W. W. Norton \& Company}, \bibinfo{address}{New York}, \bibinfo{pages}{211--245}.
\newblock
\showISBNx{978-0-393-25385-6}


\bibitem[Kaminski(2020)]%
        {kaminski2020}
\bibfield{author}{\bibinfo{person}{Margot~E. Kaminski}.} \bibinfo{year}{2020}\natexlab{}.
\newblock \showarticletitle{Understanding transparency in algorithmic accountability}.
\newblock In \bibinfo{booktitle}{\emph{Cambridge Handbook of the Law of Algorithms}}, \bibfield{editor}{\bibinfo{person}{Woodrow Barfield}} (Ed.). \bibinfo{publisher}{Cambridge University Press}, \bibinfo{pages}{20--34}.
\newblock


\bibitem[Kaur et~al\mbox{.}(2022)]%
        {kaur2022trustworthy}
\bibfield{author}{\bibinfo{person}{Davinder Kaur}, \bibinfo{person}{Suleyman Uslu}, \bibinfo{person}{Kaley~J Rittichier}, {and} \bibinfo{person}{Arjan Durresi}.} \bibinfo{year}{2022}\natexlab{}.
\newblock \showarticletitle{Trustworthy artificial intelligence: a review}.
\newblock \bibinfo{journal}{\emph{ACM Computing Surveys (CSUR)}} \bibinfo{volume}{55}, \bibinfo{number}{2} (\bibinfo{year}{2022}), \bibinfo{pages}{1--38}.
\newblock


\bibitem[Keller and Abowd(2023)]%
        {keller2023}
\bibfield{author}{\bibinfo{person}{Sallie~Ann Keller} {and} \bibinfo{person}{John~M. Abowd}.} \bibinfo{year}{2023}\natexlab{}.
\newblock \showarticletitle{Database reconstruction does compromise confidentiality}.
\newblock \bibinfo{journal}{\emph{Proceedings of the National Academy of Sciences}} \bibinfo{volume}{120}, \bibinfo{number}{12} (\bibinfo{date}{March} \bibinfo{year}{2023}), \bibinfo{pages}{e2300976120}.
\newblock
\urldef\tempurl%
\url{https://doi.org/10.1073/pnas.2300976120}
\showDOI{\tempurl}
\newblock
\shownote{Publisher: Proceedings of the National Academy of Sciences}.


\bibitem[Kenny et~al\mbox{.}(2022)]%
        {kenny2022comment}
\bibfield{author}{\bibinfo{person}{Christopher~T Kenny}, \bibinfo{person}{Shiro Kuriwaki}, \bibinfo{person}{Cory McCartan}, \bibinfo{person}{Evan~TR Rosenman}, \bibinfo{person}{Tyler Simko}, {and} \bibinfo{person}{Kosuke Imai}.} \bibinfo{year}{2022}\natexlab{}.
\newblock \showarticletitle{Comment: The Essential Role of Policy Evaluation for the 2020 Census Disclosure Avoidance System}.
\newblock \bibinfo{journal}{\emph{arXiv preprint arXiv:2210.08383}} (\bibinfo{year}{2022}).
\newblock


\bibitem[Kenny et~al\mbox{.}(2021)]%
        {kenny2021}
\bibfield{author}{\bibinfo{person}{Christopher~T. Kenny}, \bibinfo{person}{Shiro Kuriwaki}, \bibinfo{person}{Cory McCartan}, \bibinfo{person}{Evan T.~R. Rosenman}, \bibinfo{person}{Tyler Simko}, {and} \bibinfo{person}{Kosuke Imai}.} \bibinfo{year}{2021}\natexlab{}.
\newblock \showarticletitle{The use of differential privacy for census data and its impact on redistricting: {The} case of the 2020 {U}.{S}. {Census}}.
\newblock \bibinfo{journal}{\emph{Science Advances}} \bibinfo{volume}{7}, \bibinfo{number}{41} (\bibinfo{date}{Oct.} \bibinfo{year}{2021}).
\newblock
\showISSN{2375-2548}
\urldef\tempurl%
\url{https://doi.org/10.1126/sciadv.abk3283}
\showDOI{\tempurl}


\bibitem[Kimble et~al\mbox{.}(2010)]%
        {kimble2010}
\bibfield{author}{\bibinfo{person}{Chris Kimble}, \bibinfo{person}{Corinne Grenier}, {and} \bibinfo{person}{Karine Goglio-Primard}.} \bibinfo{year}{2010}\natexlab{}.
\newblock \showarticletitle{Innovation and knowledge sharing across professional boundaries: Political interplay between boundary objects and brokers}.
\newblock \bibinfo{journal}{\emph{International Journal of Information Management}} \bibinfo{volume}{30}, \bibinfo{number}{5} (\bibinfo{year}{2010}), \bibinfo{pages}{437--444}.
\newblock


\bibitem[Kluttz et~al\mbox{.}(2022)]%
        {kluttz2022shaping}
\bibfield{author}{\bibinfo{person}{Daniel~N Kluttz}, \bibinfo{person}{Nitin Kohli}, {and} \bibinfo{person}{Deirdre~K Mulligan}.} \bibinfo{year}{2022}\natexlab{}.
\newblock \showarticletitle{Shaping our tools: Contestability as a means to promote responsible algorithmic decision making in the professions}.
\newblock \bibinfo{journal}{\emph{Ethics of Data and Analytics. Auerbach Publications}} (\bibinfo{year}{2022}), \bibinfo{pages}{420--428}.
\newblock


\bibitem[Kroll(2018)]%
        {kroll2018fallacy}
\bibfield{author}{\bibinfo{person}{Joshua~A Kroll}.} \bibinfo{year}{2018}\natexlab{}.
\newblock \showarticletitle{The fallacy of inscrutability}.
\newblock \bibinfo{journal}{\emph{Philosophical Transactions of the Royal Society A: Mathematical, Physical and Engineering Sciences}} \bibinfo{volume}{376}, \bibinfo{number}{2133} (\bibinfo{year}{2018}), \bibinfo{pages}{20180084}.
\newblock


\bibitem[Kroll(2021)]%
        {kroll2021outlining}
\bibfield{author}{\bibinfo{person}{Joshua~A Kroll}.} \bibinfo{year}{2021}\natexlab{}.
\newblock \showarticletitle{Outlining traceability: A principle for operationalizing accountability in computing systems}. In \bibinfo{booktitle}{\emph{Proceedings of the 2021 ACM Conference on Fairness, Accountability, and Transparency}}. \bibinfo{pages}{758--771}.
\newblock


\bibitem[Kroll et~al\mbox{.}(2017)]%
        {kroll2017accountable}
\bibfield{author}{\bibinfo{person}{Joshua~A Kroll}, \bibinfo{person}{Joanna Huey}, \bibinfo{person}{Solon Barocas}, \bibinfo{person}{Edward~W Felten}, \bibinfo{person}{Joel~R Reidenberg}, \bibinfo{person}{David~G Robinson}, {and} \bibinfo{person}{Harlan Yu}.} \bibinfo{year}{2017}\natexlab{}.
\newblock \showarticletitle{Accountable Algorithms}.
\newblock \bibinfo{journal}{\emph{University of Pennsylvania Law Review}} \bibinfo{volume}{165}, \bibinfo{number}{3} (\bibinfo{year}{2017}), \bibinfo{pages}{633}.
\newblock


\bibitem[Langley et~al\mbox{.}(2019)]%
        {langley2019boundary}
\bibfield{author}{\bibinfo{person}{Ann Langley}, \bibinfo{person}{Kajsa Lindberg}, \bibinfo{person}{BjÃžrn~Erik MÃžrk}, \bibinfo{person}{Davide Nicolini}, \bibinfo{person}{Elena Raviola}, {and} \bibinfo{person}{Lars Walter}.} \bibinfo{year}{2019}\natexlab{}.
\newblock \showarticletitle{Boundary {Work} among {Groups}, {Occupations}, and {Organizations}: {From} {Cartography} to {Process}}.
\newblock \bibinfo{journal}{\emph{Academy of Management Annals}} \bibinfo{volume}{13}, \bibinfo{number}{2} (\bibinfo{date}{July} \bibinfo{year}{2019}), \bibinfo{pages}{704--736}.
\newblock
\showISSN{1941-6520}
\urldef\tempurl%
\url{https://doi.org/10.5465/annals.2017.0089}
\showDOI{\tempurl}
\newblock
\shownote{Publisher: Academy of Management}.


\bibitem[Lepri et~al\mbox{.}(2018)]%
        {lepri2018fair}
\bibfield{author}{\bibinfo{person}{Bruno Lepri}, \bibinfo{person}{Nuria Oliver}, \bibinfo{person}{Emmanuel Letouz{\'e}}, \bibinfo{person}{Alex Pentland}, {and} \bibinfo{person}{Patrick Vinck}.} \bibinfo{year}{2018}\natexlab{}.
\newblock \showarticletitle{Fair, transparent, and accountable algorithmic decision-making processes: The premise, the proposed solutions, and the open challenges}.
\newblock \bibinfo{journal}{\emph{Philosophy \& Technology}}  \bibinfo{volume}{31} (\bibinfo{year}{2018}), \bibinfo{pages}{611--627}.
\newblock


\bibitem[Leveson(2016)]%
        {leveson2016engineering}
\bibfield{author}{\bibinfo{person}{Nancy~G Leveson}.} \bibinfo{year}{2016}\natexlab{}.
\newblock \bibinfo{booktitle}{\emph{Engineering a safer world: Systems thinking applied to safety}}.
\newblock \bibinfo{publisher}{The MIT Press}.
\newblock


\bibitem[Levy et~al\mbox{.}(2021)]%
        {levy2021algorithms}
\bibfield{author}{\bibinfo{person}{Karen Levy}, \bibinfo{person}{Kyla~E Chasalow}, {and} \bibinfo{person}{Sarah Riley}.} \bibinfo{year}{2021}\natexlab{}.
\newblock \showarticletitle{Algorithms and decision-making in the public sector}.
\newblock \bibinfo{journal}{\emph{Annual Review of Law and Social Science}}  \bibinfo{volume}{17} (\bibinfo{year}{2021}), \bibinfo{pages}{309--334}.
\newblock


\bibitem[Lima et~al\mbox{.}(2022)]%
        {lima2022conflict}
\bibfield{author}{\bibinfo{person}{Gabriel Lima}, \bibinfo{person}{Nina Grgi{\'c}-Hla{\v{c}}a}, \bibinfo{person}{Jin~Keun Jeong}, {and} \bibinfo{person}{Meeyoung Cha}.} \bibinfo{year}{2022}\natexlab{}.
\newblock \showarticletitle{The conflict between explainable and accountable decision-making algorithms}. In \bibinfo{booktitle}{\emph{Proceedings of the 2022 ACM Conference on Fairness, Accountability, and Transparency}}. \bibinfo{pages}{2103--2113}.
\newblock


\bibitem[Loi et~al\mbox{.}(2021)]%
        {loi2021transparency}
\bibfield{author}{\bibinfo{person}{Michele Loi}, \bibinfo{person}{Andrea Ferrario}, {and} \bibinfo{person}{Eleonora Vigan{\`o}}.} \bibinfo{year}{2021}\natexlab{}.
\newblock \showarticletitle{Transparency as design publicity: explaining and justifying inscrutable algorithms}.
\newblock \bibinfo{journal}{\emph{Ethics and Information Technology}} \bibinfo{volume}{23}, \bibinfo{number}{3} (\bibinfo{year}{2021}), \bibinfo{pages}{253--263}.
\newblock


\bibitem[Madaio et~al\mbox{.}(2022)]%
        {madaio2022assessing}
\bibfield{author}{\bibinfo{person}{Michael Madaio}, \bibinfo{person}{Lisa Egede}, \bibinfo{person}{Hariharan Subramonyam}, \bibinfo{person}{Jennifer Wortman~Vaughan}, {and} \bibinfo{person}{Hanna Wallach}.} \bibinfo{year}{2022}\natexlab{}.
\newblock \showarticletitle{Assessing the Fairness of AI Systems: AI Practitioners' Processes, Challenges, and Needs for Support}.
\newblock \bibinfo{journal}{\emph{Proceedings of the ACM on Human-Computer Interaction}} \bibinfo{volume}{6}, \bibinfo{number}{CSCW1} (\bibinfo{year}{2022}), \bibinfo{pages}{1--26}.
\newblock


\bibitem[Martin(2019)]%
        {martin2019ethical}
\bibfield{author}{\bibinfo{person}{Kirsten Martin}.} \bibinfo{year}{2019}\natexlab{}.
\newblock \showarticletitle{Ethical implications and accountability of algorithms}.
\newblock \bibinfo{journal}{\emph{Journal of Business Ethics}}  \bibinfo{volume}{160} (\bibinfo{year}{2019}), \bibinfo{pages}{835--850}.
\newblock


\bibitem[McKenna(2018)]%
        {mckenna2018}
\bibfield{author}{\bibinfo{person}{Laura McKenna}.} \bibinfo{year}{2018}\natexlab{}.
\newblock \bibinfo{booktitle}{\emph{Disclosure {Avoidance} {Techniques} {Used} for the 1970 through 2010 {Decennial} {Censuses} of {Population} and {Housing}}}.
\newblock \bibinfo{type}{{T}echnical {R}eport}. \bibinfo{institution}{U.S. Census Bureau Research \& Methodology Directorate}.
\newblock
\urldef\tempurl%
\url{https://www2.census.gov/ces/wp/2018/CES-WP-18-47.pdf}
\showURL{%
\tempurl}


\bibitem[Metcalf et~al\mbox{.}(2019)]%
        {metcalf2019owning}
\bibfield{author}{\bibinfo{person}{Jacob Metcalf}, \bibinfo{person}{Emanuel Moss}, {and} \bibinfo{person}{{danah} {boyd}}.} \bibinfo{year}{2019}\natexlab{}.
\newblock \showarticletitle{Owning ethics: Corporate logics, silicon valley, and the institutionalization of ethics}.
\newblock \bibinfo{journal}{\emph{Social Research: An International Quarterly}} \bibinfo{volume}{86}, \bibinfo{number}{2} (\bibinfo{year}{2019}), \bibinfo{pages}{449--476}.
\newblock


\bibitem[{minutephysics}(2019)]%
        {minutephysics2019}
\bibfield{author}{\bibinfo{person}{{minutephysics}}.} \bibinfo{year}{2019}\natexlab{}.
\newblock \bibinfo{title}{Protecting {Privacy} with {MATH} ({Collab} with the {Census})}.
\newblock
\newblock
\urldef\tempurl%
\url{https://www.youtube.com/watch?v=pT19VwBAqKA}
\showURL{%
\tempurl}


\bibitem[Mitchell et~al\mbox{.}(2019)]%
        {mitchell2019model}
\bibfield{author}{\bibinfo{person}{Margaret Mitchell}, \bibinfo{person}{Simone Wu}, \bibinfo{person}{Andrew Zaldivar}, \bibinfo{person}{Parker Barnes}, \bibinfo{person}{Lucy Vasserman}, \bibinfo{person}{Ben Hutchinson}, \bibinfo{person}{Elena Spitzer}, \bibinfo{person}{Inioluwa~Deborah Raji}, {and} \bibinfo{person}{Timnit Gebru}.} \bibinfo{year}{2019}\natexlab{}.
\newblock \showarticletitle{Model {Cards} for {Model} {Reporting}}. In \bibinfo{booktitle}{\emph{Proceedings of the {Conference} on {Fairness}, {Accountability}, and {Transparency}}} \emph{(\bibinfo{series}{{FAT}* '19})}. \bibinfo{publisher}{Association for Computing Machinery}, \bibinfo{address}{New York, NY, USA}, \bibinfo{pages}{220--229}.
\newblock
\showISBNx{978-1-4503-6125-5}
\urldef\tempurl%
\url{https://doi.org/10.1145/3287560.3287596}
\showDOI{\tempurl}


\bibitem[Mulligan and Bamberger(2018)]%
        {mulligan2018saving}
\bibfield{author}{\bibinfo{person}{Deirdre~K Mulligan} {and} \bibinfo{person}{Kenneth~A Bamberger}.} \bibinfo{year}{2018}\natexlab{}.
\newblock \showarticletitle{Saving governance-by-design}.
\newblock \bibinfo{journal}{\emph{California Law Review}} \bibinfo{volume}{106}, \bibinfo{number}{3} (\bibinfo{year}{2018}), \bibinfo{pages}{697--784}.
\newblock


\bibitem[Mulligan and Bamberger(2019)]%
        {mulligan2019}
\bibfield{author}{\bibinfo{person}{Deirdre~K. Mulligan} {and} \bibinfo{person}{Kenneth~A. Bamberger}.} \bibinfo{year}{2019}\natexlab{}.
\newblock \showarticletitle{Procurement as policy: Administrative process for machine learning}.
\newblock \bibinfo{journal}{\emph{Berkeley Tech. LJ}}  \bibinfo{volume}{34} (\bibinfo{year}{2019}), \bibinfo{pages}{773}.
\newblock


\bibitem[Mulligan and Nissenbaum(2020)]%
        {mulligan2020}
\bibfield{author}{\bibinfo{person}{Deirdre~K. Mulligan} {and} \bibinfo{person}{Helen Nissenbaum}.} \bibinfo{year}{2020}\natexlab{}.
\newblock \showarticletitle{The concept of handoff as a model for ethical analysis and design}.
\newblock \bibinfo{journal}{\emph{The Oxford handbook of ethics of AI}} \bibinfo{volume}{1}, \bibinfo{number}{1} (\bibinfo{year}{2020}), \bibinfo{pages}{233}.
\newblock


\bibitem[Nanayakkara and Hullman(2022)]%
        {nanayakkara2022}
\bibfield{author}{\bibinfo{person}{Priyanka Nanayakkara} {and} \bibinfo{person}{Jessica Hullman}.} \bibinfo{year}{2022}\natexlab{}.
\newblock \showarticletitle{What's driving conflicts around differential privacy for the U.S. Census}.
\newblock \bibinfo{journal}{\emph{IEEE Security \& Privacy}} \bibinfo{number}{01} (\bibinfo{year}{2022}), \bibinfo{pages}{2--11}.
\newblock


\bibitem[{National Academies of Sciences, Engineering, and Medicine}(2020)]%
        {nasem2020}
\bibfield{author}{\bibinfo{person}{{National Academies of Sciences, Engineering, and Medicine}}.} \bibinfo{year}{2020}\natexlab{}.
\newblock \bibinfo{booktitle}{\emph{2020 Census Data Products: Data Needs and Privacy Considerations: Proceedings of a Workshop}}.
\newblock \bibinfo{publisher}{National Academies Press}.
\newblock


\bibitem[{National Conference of State Legislatures}(2021)]%
        {ncsl21}
\bibfield{author}{\bibinfo{person}{{National Conference of State Legislatures}}.} \bibinfo{year}{2021}\natexlab{}.
\newblock \bibinfo{title}{Differential {Privacy} for {Census} {Data} {Explained}}.
\newblock
\newblock
\urldef\tempurl%
\url{https://www.ncsl.org/technology-and-communication/differential-privacy-for-census-data-explained}
\showURL{%
\tempurl}


\bibitem[{NCAI Policy Research Center}(2021)]%
        {ncai2021}
\bibfield{author}{\bibinfo{person}{{NCAI Policy Research Center}}.} \bibinfo{year}{2021}\natexlab{}.
\newblock \bibinfo{booktitle}{\emph{Differential {Privacy} and the 2020 {Census}: {A} {Guide} to the {Data} {Analyses} and {Impacts} on {AI}/{AN} {Data}}}.
\newblock \bibinfo{type}{Research {Policy} {Update}}. \bibinfo{institution}{National Congress of American Indians}, \bibinfo{address}{Washington, D.C.}
\newblock
\urldef\tempurl%
\url{https://archive.ncai.org/policy-research-center/research-data/prc-publications/NCAI_PRC_2020_Census_Guide_to_Data_and_Impacts_5_17_2021_FINAL.pdf}
\showURL{%
\tempurl}


\bibitem[Nissenbaum(2004)]%
        {nissenbaum2004privacy}
\bibfield{author}{\bibinfo{person}{Helen Nissenbaum}.} \bibinfo{year}{2004}\natexlab{}.
\newblock \showarticletitle{Privacy as contextual integrity}.
\newblock \bibinfo{journal}{\emph{Wash. L. Rev.}}  \bibinfo{volume}{79} (\bibinfo{year}{2004}), \bibinfo{pages}{119}.
\newblock


\bibitem[Ochoa and Minnis(2021)]%
        {ochoa2021}
\bibfield{author}{\bibinfo{person}{Steven~A. Ochoa} {and} \bibinfo{person}{Terry~Ao Minnis}.} \bibinfo{year}{2021}\natexlab{}.
\newblock \bibinfo{title}{Impact of Differential Privacy \& the 2020 Census on Latinos, Asian Americans and Redistricting}.
\newblock
\newblock
\urldef\tempurl%
\url{https://www.maldef.org/wp-content/uploads/2021/04/FINAL-MALDEF-AAJC-Differential-Privacy-Preliminary-Report-4.5.2021-1.pdf}
\showURL{%
\tempurl}


\bibitem[OECD(2021)]%
        {OECD_2021}
\bibfield{author}{\bibinfo{person}{OECD}.} \bibinfo{year}{2021}\natexlab{}.
\newblock \showarticletitle{Tools for trustworthy AI: A framework to compare implementation tools for trustworthy AI systems}.
\newblock \bibinfo{journal}{\emph{OECD Digital Economy Papers}} \bibinfo{number}{312} (\bibinfo{date}{Jun} \bibinfo{year}{2021}).
\newblock
\urldef\tempurl%
\url{https://doi.org/10.1787/008232ec-en}
\showDOI{\tempurl}


\bibitem[of~Standards and Technology(2023)]%
        {NIST_2023}
\bibfield{author}{\bibinfo{person}{National~Institute of Standards} {and} \bibinfo{person}{Technology}.} \bibinfo{year}{2023}\natexlab{}.
\newblock \showarticletitle{Artificial Intelligence Risk Management Framework (AI RMF 1.0)}.
\newblock  (\bibinfo{date}{Jan} \bibinfo{year}{2023}).
\newblock
\urldef\tempurl%
\url{https://doi.org/10.6028/nist.ai.100-1}
\showDOI{\tempurl}


\bibitem[Pierre et~al\mbox{.}(2021)]%
        {pierre2021getting}
\bibfield{author}{\bibinfo{person}{Jennifer Pierre}, \bibinfo{person}{Roderic Crooks}, \bibinfo{person}{Morgan Currie}, \bibinfo{person}{Britt Paris}, {and} \bibinfo{person}{Irene Pasquetto}.} \bibinfo{year}{2021}\natexlab{}.
\newblock \showarticletitle{Getting Ourselves Together: Data-centered participatory design research \& epistemic burden}. In \bibinfo{booktitle}{\emph{Proceedings of the 2021 CHI Conference on Human Factors in Computing Systems}}. \bibinfo{pages}{1--11}.
\newblock


\bibitem[Porter(1995)]%
        {porter1995}
\bibfield{author}{\bibinfo{person}{Theodore~M. Porter}.} \bibinfo{year}{1995}\natexlab{}.
\newblock \bibinfo{booktitle}{\emph{Trust in numbers: The pursuit of objectivity in science and public life}}.
\newblock \bibinfo{publisher}{Princeton University Press}.
\newblock


\bibitem[Raji et~al\mbox{.}(2020)]%
        {raji2020closing}
\bibfield{author}{\bibinfo{person}{Inioluwa~Deborah Raji}, \bibinfo{person}{Andrew Smart}, \bibinfo{person}{Rebecca~N White}, \bibinfo{person}{Margaret Mitchell}, \bibinfo{person}{Timnit Gebru}, \bibinfo{person}{Ben Hutchinson}, \bibinfo{person}{Jamila Smith-Loud}, \bibinfo{person}{Daniel Theron}, {and} \bibinfo{person}{Parker Barnes}.} \bibinfo{year}{2020}\natexlab{}.
\newblock \showarticletitle{Closing the AI accountability gap: Defining an end-to-end framework for internal algorithmic auditing}. In \bibinfo{booktitle}{\emph{Proceedings of the 2020 conference on fairness, accountability, and transparency}}. \bibinfo{pages}{33--44}.
\newblock


\bibitem[Rawlins(2008)]%
        {rawlins2008}
\bibfield{author}{\bibinfo{person}{Brad Rawlins}.} \bibinfo{year}{2008}\natexlab{}.
\newblock \showarticletitle{Give the emperor a mirror: Toward developing a stakeholder measurement of organizational transparency}.
\newblock \bibinfo{journal}{\emph{Journal of public relations research}} \bibinfo{volume}{21}, \bibinfo{number}{1} (\bibinfo{year}{2008}), \bibinfo{pages}{71--99}.
\newblock


\bibitem[Robinson(2022)]%
        {robinson2022voices}
\bibfield{author}{\bibinfo{person}{David~Gerald Robinson}.} \bibinfo{year}{2022}\natexlab{}.
\newblock \bibinfo{booktitle}{\emph{Voices in the code: a story about people, their values, and the algorithm they made}}.
\newblock \bibinfo{publisher}{Russell Sage Foundation}, \bibinfo{address}{New York}.
\newblock
\showISBNx{978-0-87154-777-4}


\bibitem[Sarkki et~al\mbox{.}(2020)]%
        {sarkki2020}
\bibfield{author}{\bibinfo{person}{Simo Sarkki}, \bibinfo{person}{Hannu~I. Heikkinen}, \bibinfo{person}{Teresa Komu}, \bibinfo{person}{Mari Partanen}, \bibinfo{person}{Karoliina Vanhanen}, {and} \bibinfo{person}{Elise Lepy}.} \bibinfo{year}{2020}\natexlab{}.
\newblock \showarticletitle{How boundary objects help to perform roles of science arbiter, honest broker, and issue advocate}.
\newblock \bibinfo{journal}{\emph{Science and Public Policy}} \bibinfo{volume}{47}, \bibinfo{number}{2} (\bibinfo{year}{2020}), \bibinfo{pages}{161--171}.
\newblock


\bibitem[Schnackenberg and Tomlinson(2016)]%
        {schnackenberg2016}
\bibfield{author}{\bibinfo{person}{Andrew~K. Schnackenberg} {and} \bibinfo{person}{Edward~C. Tomlinson}.} \bibinfo{year}{2016}\natexlab{}.
\newblock \showarticletitle{Organizational transparency: A new perspective on managing trust in organization-stakeholder relationships}.
\newblock \bibinfo{journal}{\emph{Journal of management}} \bibinfo{volume}{42}, \bibinfo{number}{7} (\bibinfo{year}{2016}), \bibinfo{pages}{1784--1810}.
\newblock


\bibitem[Schneider(2021)]%
        {schneider2021}
\bibfield{author}{\bibinfo{person}{Mike Schneider}.} \bibinfo{year}{2021}\natexlab{}.
\newblock \bibinfo{title}{Census releases guidelines for controversial privacy tool}.
\newblock
\newblock
\urldef\tempurl%
\url{https://apnews.com/article/business-census-2020-55519b7534bd8d61028020d79854e909}
\showURL{%
\tempurl}
\newblock
\shownote{Section: Voting rights}.


\bibitem[Seeman(2023)]%
        {seeman2023framing}
\bibfield{author}{\bibinfo{person}{Jeremy Seeman}.} \bibinfo{year}{2023}\natexlab{}.
\newblock \showarticletitle{Framing Effects in the Operationalization of Differential Privacy Systems as Code-Driven Law}. In \bibinfo{booktitle}{\emph{International Conference on Computer Ethics}}, Vol.~\bibinfo{volume}{1}.
\newblock


\bibitem[Seeman and Susser(2023)]%
        {seeman2023between}
\bibfield{author}{\bibinfo{person}{Jeremy Seeman} {and} \bibinfo{person}{Daniel Susser}.} \bibinfo{year}{2023}\natexlab{}.
\newblock \showarticletitle{Between Privacy and Utility: On Differential Privacy in Theory and Practice}.
\newblock \bibinfo{journal}{\emph{ACM J. Responsib. Comput.}} (\bibinfo{date}{oct} \bibinfo{year}{2023}).
\newblock
\urldef\tempurl%
\url{https://doi.org/10.1145/3626494}
\showDOI{\tempurl}
\newblock
\shownote{Just Accepted}.


\bibitem[Selbst et~al\mbox{.}(2019)]%
        {selbst2019fairness}
\bibfield{author}{\bibinfo{person}{Andrew~D Selbst}, \bibinfo{person}{Danah Boyd}, \bibinfo{person}{Sorelle~A Friedler}, \bibinfo{person}{Suresh Venkatasubramanian}, {and} \bibinfo{person}{Janet Vertesi}.} \bibinfo{year}{2019}\natexlab{}.
\newblock \showarticletitle{Fairness and abstraction in sociotechnical systems}. In \bibinfo{booktitle}{\emph{Proceedings of the conference on fairness, accountability, and transparency}}. \bibinfo{pages}{59--68}.
\newblock


\bibitem[Sloane and Moss(2022)]%
        {sloane2022introducing}
\bibfield{author}{\bibinfo{person}{Mona Sloane} {and} \bibinfo{person}{Emanuel Moss}.} \bibinfo{year}{2022}\natexlab{}.
\newblock \showarticletitle{Introducing a Practice-Based Compliance Framework for Addressing New Regulatory Challenges in the AI Field}.
\newblock \bibinfo{journal}{\emph{TechReg Chronicle}} (\bibinfo{year}{2022}).
\newblock


\bibitem[Sloane et~al\mbox{.}(2022)]%
        {sloane2022}
\bibfield{author}{\bibinfo{person}{Mona Sloane}, \bibinfo{person}{Emanuel Moss}, \bibinfo{person}{Olaitan Awomolo}, {and} \bibinfo{person}{Laura Forlano}.} \bibinfo{year}{2022}\natexlab{}.
\newblock \showarticletitle{Participation is not a design fix for machine learning}.
\newblock In \bibinfo{booktitle}{\emph{Equity and Access in Algorithms, Mechanisms, and Optimization}}. \bibinfo{pages}{1--6}.
\newblock


\bibitem[Sloane et~al\mbox{.}(2023)]%
        {sloane2023introducingcontextual}
\bibfield{author}{\bibinfo{person}{Mona Sloane}, \bibinfo{person}{Ian~Ren{\'e} Solano-Kamaiko}, \bibinfo{person}{Jun Yuan}, \bibinfo{person}{Aritra Dasgupta}, {and} \bibinfo{person}{Julia Stoyanovich}.} \bibinfo{year}{2023}\natexlab{}.
\newblock \showarticletitle{Introducing contextual transparency for automated decision systems}.
\newblock \bibinfo{journal}{\emph{Nature Machine Intelligence}} \bibinfo{volume}{5}, \bibinfo{number}{3} (\bibinfo{year}{2023}), \bibinfo{pages}{187--195}.
\newblock


\bibitem[Sloane and Zakrzewski(2022)]%
        {sloane2022german}
\bibfield{author}{\bibinfo{person}{Mona Sloane} {and} \bibinfo{person}{Janina Zakrzewski}.} \bibinfo{year}{2022}\natexlab{}.
\newblock \showarticletitle{German AI Start-Ups and “AI Ethics”: Using A Social Practice Lens for Assessing and Implementing Socio-Technical Innovation}. In \bibinfo{booktitle}{\emph{Proceedings of the 2022 ACM Conference on Fairness, Accountability, and Transparency}}. \bibinfo{pages}{935--947}.
\newblock


\bibitem[Star(2010)]%
        {star2010}
\bibfield{author}{\bibinfo{person}{Susan~Leigh Star}.} \bibinfo{year}{2010}\natexlab{}.
\newblock \showarticletitle{This is not a boundary object: Reflections on the origin of a concept}.
\newblock \bibinfo{journal}{\emph{Science, Technology, \& Human Values}} \bibinfo{volume}{35}, \bibinfo{number}{5} (\bibinfo{year}{2010}), \bibinfo{pages}{601--617}.
\newblock


\bibitem[Star and Griesemer(1989)]%
        {star1989}
\bibfield{author}{\bibinfo{person}{Susan~Leigh Star} {and} \bibinfo{person}{James~R. Griesemer}.} \bibinfo{year}{1989}\natexlab{}.
\newblock \showarticletitle{Institutional ecology,translations' and boundary objects: Amateurs and professionals in Berkeley's Museum of Vertebrate Zoology, 1907-39}.
\newblock \bibinfo{journal}{\emph{Social Studies of Science}} \bibinfo{volume}{19}, \bibinfo{number}{3} (\bibinfo{year}{1989}), \bibinfo{pages}{387--420}.
\newblock


\bibitem[Steed et~al\mbox{.}(2022)]%
        {steed2022}
\bibfield{author}{\bibinfo{person}{Ryan Steed}, \bibinfo{person}{Terrance Liu}, \bibinfo{person}{Zhiwei~Steven Wu}, {and} \bibinfo{person}{Alessandro Acquisti}.} \bibinfo{year}{2022}\natexlab{}.
\newblock \showarticletitle{Policy impacts of statistical uncertainty and privacy}.
\newblock \bibinfo{journal}{\emph{Science}} \bibinfo{volume}{377}, \bibinfo{number}{6609} (\bibinfo{year}{2022}), \bibinfo{pages}{928--931}.
\newblock
\urldef\tempurl%
\url{https://doi.org/10.1126/science.abq4481}
\showDOI{\tempurl}
\showeprint{https://www.science.org/doi/pdf/10.1126/science.abq4481}


\bibitem[Sunstein(2002)]%
        {sunstein2002}
\bibfield{author}{\bibinfo{person}{Cass~R Sunstein}.} \bibinfo{year}{2002}\natexlab{}.
\newblock \showarticletitle{The cost-benefit state: the future of regulatory protection}. American Bar Association.
\newblock


\bibitem[Sweeney(2000)]%
        {sweeney2000}
\bibfield{author}{\bibinfo{person}{Latanya Sweeney}.} \bibinfo{year}{2000}\natexlab{}.
\newblock \bibinfo{booktitle}{\emph{Simple Demographics Often Identify People Uniquely}}.
\newblock \bibinfo{type}{Working Paper}. \bibinfo{institution}{Carnegie Mellon University Data Privacy Lab}, \bibinfo{address}{Pittsburgh}.
\newblock
\urldef\tempurl%
\url{https://dataprivacylab.org/projects/identifiability/paper1.pdf}
\showURL{%
\tempurl}


\bibitem[{U.S. Census Bureau}(2018)]%
        {fedregister2018}
\bibfield{author}{\bibinfo{person}{{U.S. Census Bureau}}.} \bibinfo{year}{2018}\natexlab{}.
\newblock \bibinfo{title}{Soliciting {Feedback} {From} {Users} on 2020 {Census} {Data} {Products}}.
\newblock
\newblock
\urldef\tempurl%
\url{https://www.federalregister.gov/documents/2018/07/19/2018-15458/soliciting-feedback-from-users-on-2020-census-data-products}
\showURL{%
\tempurl}


\bibitem[{U.S. Census Bureau}(2020a)]%
        {consults2020}
\bibfield{author}{\bibinfo{person}{{U.S. Census Bureau}}.} \bibinfo{year}{2020}\natexlab{a}.
\newblock \bibinfo{booktitle}{\emph{2020 {Census} {Tribal} {Consultations} with {Federally} {Recognized} {Tribes}}}.
\newblock \bibinfo{type}{Report}. \bibinfo{institution}{U.S. Census Bureau}.
\newblock
\urldef\tempurl%
\url{https://www.census.gov/content/dam/Census/library/publications/2020/dec/census-federal-tc-final-report-2020-508.pdf}
\showURL{%
\tempurl}


\bibitem[{U.S. Census Bureau}(2020b)]%
        {cbinvariants2020}
\bibfield{author}{\bibinfo{person}{{U.S. Census Bureau}}.} \bibinfo{year}{2020}\natexlab{b}.
\newblock \bibinfo{title}{Invariants {Set} for 2020 {Census} {Data} {Products}}.
\newblock
\newblock
\urldef\tempurl%
\url{https://www.census.gov/programs-surveys/decennial-census/decade/2020/planning-management/process/disclosure-avoidance/2020-das-updates/2020-11-25.html}
\showURL{%
\tempurl}


\bibitem[{U.S. Census Bureau}(2021a)]%
        {epsilon2021}
\bibfield{author}{\bibinfo{person}{{U.S. Census Bureau}}.} \bibinfo{year}{2021}\natexlab{a}.
\newblock \bibinfo{title}{Census {Bureau} {Sets} {Key} {Parameters} to {Protect} {Privacy} in 2020 {Census} {Results}}.
\newblock
\newblock
\urldef\tempurl%
\url{https://www.census.gov/newsroom/press-releases/2021/2020-census-key-parameters.html}
\showURL{%
\tempurl}
\newblock
\shownote{Section: Government}.


\bibitem[{U.S. Census Bureau}(2021b)]%
        {cb2021}
\bibfield{author}{\bibinfo{person}{{U.S. Census Bureau}}.} \bibinfo{year}{2021}\natexlab{b}.
\newblock \bibinfo{booktitle}{\emph{Disclosure {Avoidance} for the 2020 {Census}: {An} {Introduction}}}.
\newblock \bibinfo{type}{Handbook}. \bibinfo{institution}{US Government Publishing Office}, \bibinfo{address}{Washington, D.C.}
\newblock
\urldef\tempurl%
\url{https://www2.census.gov/library/publications/decennial/2020/2020-census-disclosure-avoidance-handbook.pdf}
\showURL{%
\tempurl}


\bibitem[{U.S. Census Bureau}(2023a)]%
        {bureau2023}
\bibfield{author}{\bibinfo{person}{{U.S. Census Bureau}}.} \bibinfo{year}{2023}\natexlab{a}.
\newblock \bibinfo{title}{2020 Decennial Census: Processing the Count: Disclosure Avoidance Modernization}.
\newblock
\newblock
\urldef\tempurl%
\url{https://www.census.gov/programs-surveys/decennial-census/decade/2020/planning-management/process/disclosure-avoidance.html}
\showURL{%
\tempurl}


\bibitem[{U.S. Census Bureau}(2023b)]%
        {cb2023}
\bibfield{author}{\bibinfo{person}{{U.S. Census Bureau}}.} \bibinfo{year}{2023}\natexlab{b}.
\newblock \bibinfo{title}{Coming {This} {Spring}: {New} 2010 {Redistricting} and {DHC} "{Production} {Settings}" {Demonstration} {Microdata} with {Noisy} {Measurement} {Files}}.
\newblock
\newblock
\urldef\tempurl%
\url{https://www.census.gov/programs-surveys/decennial-census/decade/2020/planning-management/process/disclosure-avoidance/newsletters/new-2010-redistricting-dhc-demo-microdata.html}
\showURL{%
\tempurl}
\newblock
\shownote{Section: Government}.


\bibitem[{U.S. Census Bureau}(2023c)]%
        {webinar2023}
\bibfield{author}{\bibinfo{person}{{U.S. Census Bureau}}.} \bibinfo{year}{2023}\natexlab{c}.
\newblock \bibinfo{title}{Disclosure {Avoidance} {Webinar} {Series}}.
\newblock
\newblock
\urldef\tempurl%
\url{https://www.census.gov/data/academy/webinars/series/disclosure-avoidance.html}
\showURL{%
\tempurl}
\newblock
\shownote{Section: Government}.


\bibitem[{U.S. Census Bureau}(2023d)]%
        {bureau2023why}
\bibfield{author}{\bibinfo{person}{{U.S. Census Bureau}}.} \bibinfo{year}{2023}\natexlab{d}.
\newblock \bibinfo{booktitle}{\emph{Why the Census Bureau Chose Differential Privacy}}.
\newblock \bibinfo{type}{Brief} C2020BR-03. \bibinfo{institution}{U.S. Census Bureau}.
\newblock
\urldef\tempurl%
\url{https://www2.census.gov/library/publications/decennial/2020/census-briefs/c2020br-03.pdf}
\showURL{%
\tempurl}


\bibitem[Vaccaro et~al\mbox{.}(2019)]%
        {vaccaro2019contestability}
\bibfield{author}{\bibinfo{person}{Kristen Vaccaro}, \bibinfo{person}{Karrie Karahalios}, \bibinfo{person}{Deirdre~K Mulligan}, \bibinfo{person}{Daniel Kluttz}, {and} \bibinfo{person}{Tad Hirsch}.} \bibinfo{year}{2019}\natexlab{}.
\newblock \showarticletitle{Contestability in algorithmic systems}. In \bibinfo{booktitle}{\emph{Conference companion publication of the 2019 on computer supported cooperative work and social computing}}. \bibinfo{pages}{523--527}.
\newblock


\bibitem[Van~Riper et~al\mbox{.}(2021)]%
        {vanriper2021}
\bibfield{author}{\bibinfo{person}{David Van~Riper}, \bibinfo{person}{Jonathan Schroeder}, {and} \bibinfo{person}{Steven Ruggles}.} \bibinfo{year}{2021}\natexlab{}.
\newblock \bibinfo{title}{Feedback on the April 2021 Census Demonstration Files}.
\newblock
\newblock
\urldef\tempurl%
\url{https://users.pop.umn.edu/~ruggl001/Articles/IPUMS_response_to_Census.pdf}
\showURL{%
\tempurl}


\bibitem[Veale et~al\mbox{.}(2018)]%
        {veale2018fairness}
\bibfield{author}{\bibinfo{person}{Michael Veale}, \bibinfo{person}{Max Van~Kleek}, {and} \bibinfo{person}{Reuben Binns}.} \bibinfo{year}{2018}\natexlab{}.
\newblock \showarticletitle{Fairness and accountability design needs for algorithmic support in high-stakes public sector decision-making}. In \bibinfo{booktitle}{\emph{Proceedings of the 2018 chi conference on human factors in computing systems}}. \bibinfo{pages}{1--14}.
\newblock


\bibitem[Viljoen(2021)]%
        {viljoen2021relational}
\bibfield{author}{\bibinfo{person}{Salome Viljoen}.} \bibinfo{year}{2021}\natexlab{}.
\newblock \showarticletitle{A relational theory of data governance}.
\newblock \bibinfo{journal}{\emph{Yale Law Journal}}  \bibinfo{volume}{131} (\bibinfo{year}{2021}), \bibinfo{pages}{573}.
\newblock


\bibitem[Wieringa(2020)]%
        {wieringa2020account}
\bibfield{author}{\bibinfo{person}{Maranke Wieringa}.} \bibinfo{year}{2020}\natexlab{}.
\newblock \showarticletitle{What to account for when accounting for algorithms: a systematic literature review on algorithmic accountability}. In \bibinfo{booktitle}{\emph{Proceedings of the 2020 conference on fairness, accountability, and transparency}}. \bibinfo{pages}{1--18}.
\newblock


\bibitem[Williams and Bowen(2023)]%
        {williams2023}
\bibfield{author}{\bibinfo{person}{Aaron~R. Williams} {and} \bibinfo{person}{Claire~McKay Bowen}.} \bibinfo{year}{2023}\natexlab{}.
\newblock \showarticletitle{The promise and limitations of formal privacy}.
\newblock \bibinfo{journal}{\emph{Wiley Interdisciplinary Reviews: Computational Statistics}} (\bibinfo{year}{2023}), \bibinfo{pages}{e1615}.
\newblock


\bibitem[Wong(2020)]%
        {wong2020}
\bibfield{author}{\bibinfo{person}{Richmond~Y. Wong}.} \bibinfo{year}{2020}\natexlab{}.
\newblock \emph{\bibinfo{title}{Values by {Design} {Imaginaries}: {Exploring} {Values} {Work} in {UX} {Practice}}}.
\newblock {PhD} {Dissertation}. \bibinfo{school}{University of California, Berkeley}, \bibinfo{address}{Berkeley, California}.
\newblock


\bibitem[Wong et~al\mbox{.}(2023)]%
        {wong2023seeing}
\bibfield{author}{\bibinfo{person}{Richmond~Y Wong}, \bibinfo{person}{Michael~A Madaio}, {and} \bibinfo{person}{Nick Merrill}.} \bibinfo{year}{2023}\natexlab{}.
\newblock \showarticletitle{Seeing like a toolkit: How toolkits envision the work of AI ethics}.
\newblock \bibinfo{journal}{\emph{Proceedings of the ACM on Human-Computer Interaction}} \bibinfo{volume}{7}, \bibinfo{number}{CSCW1} (\bibinfo{year}{2023}), \bibinfo{pages}{1--27}.
\newblock


\bibitem[Wright(2022)]%
        {wright2022}
\bibfield{author}{\bibinfo{person}{Larry Wright, Jr.}} \bibinfo{year}{2022}\natexlab{}.
\newblock \bibinfo{title}{Letter from {National} {Congress} of {American} {Indians} {CEO} to to {US} {Census} {Director}}.
\newblock
\newblock
\urldef\tempurl%
\url{https://www.ncai.org/policy-research-center/research-data/prc-publications/20220728_NCAI_Letter_to_US_Census_Bureau_FINAL.pdf}
\showURL{%
\tempurl}


\bibitem[Wu(2013)]%
        {wu2013}
\bibfield{author}{\bibinfo{person}{Felix~T Wu}.} \bibinfo{year}{2013}\natexlab{}.
\newblock \showarticletitle{Defining {Privacy} and {Utility} in {Data} {Sets}}.
\newblock \bibinfo{journal}{\emph{University of Colorado Law Review}}  \bibinfo{volume}{84} (\bibinfo{year}{2013}), \bibinfo{pages}{1117--1177}.
\newblock


\end{thebibliography}

\end{document}